\documentclass[10pt,aps,pra,twocolumn,longbibliography,superscriptaddress,doublespace]{revtex4-1}

\usepackage{graphicx}
\usepackage{setspace,color,appendix,physics,wasysym,xcolor,xparse}
\usepackage{amsmath,amsfonts,amssymb,mathtools,dsfont}
\usepackage{appendix}
\usepackage[english]{babel}
\usepackage{amsmath,amssymb,amsfonts,mathtools}
\usepackage{bm,bbm,euscript,braket,esint}
\definecolor{brickred}{rgb}{0.8, 0.25, 0.33}
\definecolor{celestialblue}{rgb}{0.29, 0.59, 0.82}
\definecolor{cornflowerblue}{rgb}{0.39, 0.58, 0.93}
\definecolor{denim}{rgb}{0.08, 0.38, 0.74}
\definecolor{armygreen}{rgb}{0.29, 0.33, 0.13}
\definecolor{cardinal}{rgb}{0.77, 0.12, 0.23}
\definecolor{carnelian}{rgb}{0.7, 0.11, 0.11}
\definecolor{armygreen}{rgb}{0.29, 0.33, 0.13}
\usepackage{amsmath}
\usepackage{float}

\usepackage[colorlinks=true, urlcolor=black, citecolor=blue, linkcolor=red, citebordercolor={1 0 0}, linkbordercolor={0 0 1}]{hyperref}

\newcommand\identity{1\kern-0.25em\text{l}}

\usepackage{verbatim,multirow}
\usepackage[utf8]{inputenc}
\usepackage[T1]{fontenc}
\usepackage{lmodern}
\usepackage{subcaption}

\begin{document}

\title{Characterisation of a satellite-to-ground channel for continuous variable quantum key distribution protocol}
\author{Emma Tien Hwai Medlock}
\email{ethm500@york.ac.uk}
\affiliation{School of Physics, Engineering \& Technology and York Centre for Quantum Technologies, University of York, YO10 5FT York, U.K.}
\author{Vinod N. Rao}
\affiliation{School of Physics, Engineering \& Technology and York Centre for Quantum Technologies, University of York, YO10 5FT York, U.K.}
\affiliation{Integrated Quantum Networks Hub, U.K.}
\author{Timothy Spiller}
\affiliation{School of Physics, Engineering \& Technology and York Centre for Quantum Technologies, University of York, YO10 5FT York, U.K.}
\author{Rupesh Kumar}
\email{rupesh.kumar@york.ac.uk}
\affiliation{School of Physics, Engineering \& Technology and York Centre for Quantum Technologies, University of York, YO10 5FT York, U.K.}
\affiliation{Integrated Quantum Networks Hub, U.K.}

\begin{abstract}
In space based quantum key distribution (QKD) protocols, the quantum channel will be dynamic in nature and the channel loss will change with respect to the zenith angle. In the context of continuous variable (CV)-QKD, this will cause issues with parameter estimation and for a transmitted local oscillator in particular it will also fluctuate the shot noise. Therefore, it is vital to characterise this channel loss and the sources of this loss. In this paper the varying channel loss is characterised under practical assumptions. This is shown for various different scenarios, turbulence strengths, as well as wavelengths. This work shows, for the channel parameters considered, it is possible to generate a positive secret key if restricted Eve security assumptions are made.
\end{abstract}

\maketitle

\section{Introduction}
The establishment of a secret key between remote users, Alice and Bob, via quantum key distribution (QKD) protocol utilises quantum signals for encoding the key, a quantum channel for signal transport and quantum sensitive detectors for decoding the key, along with an auxiliary classical communication channel for data reconciliation \cite{bennett1984update, bennett1987quantum, pirandola2020advances}. In contrast to classical methods, whose security relies on the computational complexity of mathematical problems \cite{bhatia2020efficient}, QKD provides information-theoretic security, provable even in the presence of a quantum-enabled adversary \cite{diamanti2015distributing, rarity2002ground}. While terrestrial fibre-based QKD systems are commercially available \cite{stanley2022recent}, their scalability is limited by losses. Even though multiple trusted relays are used, it is still difficult to reach intercontinental distances \cite{lucamarini2018overcoming}.

Space-based QKD has emerged as a potential solution to overcome these distance constraints. Since the first demonstration in 2017 \cite{lu2022micius} by the Chinese satellite Micius, there have been many other QKD satellite missions under development. This includes UK SPOQC mission \cite{SPOQC}, the Canadian QEYSsat mission \cite{scott2020qeyssat}, the European Eagel-1 mission \cite{hiemstra2025european}, the German Qube mission \cite{knips2022qube} and many more \cite{vergoossen2020spooqy, balakier2025high, jennewein2014nanoqey, oi2017cubesat, armengol2008quantum, sivasankaran2022cubesat}. Most of these missions are aimed at demonstrating a quantum communication scheme using discrete variable (DV) based protocols. These are the protocols wherein the encoding and decoding happen for a certain set of weak coherent states and thus involve an average photon per pulse $<1$. Contrary to this, continuous variables (CV) based quantum communications utilises Gaussian states for key exchange.

CV-QKD is considered particularly advantageous for space-based implementation due to the inherent noise filtering capability of its strong local oscillator (LO) in coherent detection \cite{pirandola2021limits}. The LO effectively creates a narrow spectral filter through the interferometric mixing with the QKD signal, enabling secure key distribution over noisy channels, such as those encountered in daytime satellite-to-ground QKD. With the noise filtering effect and viability of CV-QKD under free space, daytime conditions have already been experimentally demonstrated \cite{heim2014atmospheric}. However, the channel losses in space-based Low Earth Orbit (LEO) QKD are dynamic in nature compared to a static fibre-based quantum channels \cite{yehia2023connecting}. The dynamic channel loss, which depends on the variable propagation distance between satellite and ground station, arises from phenomena like beam wandering and broadening (due to atmospheric turbulence and diffraction), telescope pointing inaccuracies, atmospheric absorption, and intensity fluctuations due to scintillation \cite{vasylyev2019satellite}. Additional losses are incurred at the receiver, Bob, during coupling of the QKD signal to the detection aperture \cite{kumar2021increasing}. 

Characterisation of the losses in a free space channel for communication has been carried out over decades \cite{bouchet2010free, kaushal2017free, chan2006free, bloom2003understanding, khalighi2014survey, guiomar2022coherent}. The modelling of the atmospheric conditions for uplink- or downlink-based schemes is of utmost importance for the satellite-based classical optical communication scenarios, and one can leverage various results of those models. However, considering the atmosphere as a quantum channel, with the presence of quantum enables eavesdropper, brings various challenges in space-based quantum communications. We address some of these challenges in this manuscript and provide results that are important for the downlink-based CV-QKD optical link from the upcoming SPOQC mission.

Our study investigates the characteristics of the dynamic channel loss and its impact on the achievable key rates for the SPOQC mission CV-QKD protocol under various weather conditions. The methodology involves introducing and characterising the channel loss parameters in Sec. \ref{channel_loss}, by first defining the geometrical aspects of the satellite's pass over the Optical Ground Station (OGS), followed by the computation of relevant turbulence parameters, and finally a detailed characterisation and analysis of the various loss mechanisms. Theoretical analysis on the secure key generation of space based CV-QKD has been studied, but has not been experimentally demonstrated so far. We provide the achievable key rates for the SPOQC mission CV-QKD protocol, given the loss expected for the channel under restricted Eve security assumptions in Sec. \ref{key_rate} and finally the paper is concluded in Sec. \ref{conclusion}.

\section{Characterisation of Channel loss}
\label{channel_loss}
In the following section, the different parameters that contribute to the overall channel loss are characterised. These losses can either be geometrical losses or losses due to turbulence conditions. Beam diffraction, atmospheric attenuation and satellite pointing error belong to the former category as they depend on the propagating path length \cite{gonzalez2023satellite}, whilst beam broadening \& wandering due to turbulence and scintillation belong to the latter category. Losses arising form various weather conditions are also considered in this work. For this work the satellite parameters were chosen as those of the SPOQC mission. This includes the key pass occurring between zenith angles of $-30^\circ$ and $+30^\circ$. The key exchange could happen at angles closer to the horizon, although this will incur more loss. Here, the idea is that pointing \& tracking using uplink/downlink beacons will initialise closer to the horizon, at the beginning of the satellite pass. Only after this stage, the quantum communication commences. The classical post processing communication will occur towards the end of the satellite pass or at other available passes. The working parameters considered in this work and for the SPOQC satellite are shown in the Tab \ref{tab:SPOQC_param}. We consider a satellite trajectory aligned with the zenith angle with respect to the OGS. The channel loss will be characterised in this section for the cases of low, medium and strong turbulence conditions, with various receiver telescope apertures at the OGS. 

\begin{table}[]
\begin{center}
\begin{tabular}{||c c||} 
\hline
Parameter & Value in SPOQC mission\\ [0.5ex] 
\hline\hline
Satellite altitude & $550~\text{km}$ \\ 
\hline
Satellite telescope aperture & $8~\text{cm}$ \\
\hline
CV-QKD wavelength & $1550~\text{nm}$ \\
\hline
Pointing error & $4~\mu\text{rad}$ \\ 
\hline
Uplink beacon & $850~\text{nm}$ \\ 
\hline
Downlink beacon & $685~\text{nm}$ \\ [1ex] 
\hline
Low turbulence strength& $C_N^2(h_0)=10^{-14}~\text{m}^{-2/3}$\\ 
\hline
Medium turbulence strength& $C_N^2(h_0)=10^{-13}~\text{m}^{-2/3}$ \\
\hline
High turbulence strength & $C_N^2(h_0)=10^{-12}~\text{m}^{-2/3}$\\
\hline
Extinction coefficient $1550~\text{nm}$ & $1.8\times10^{-7}~\text{km}^{-1}$ \cite{elterman1968uv}\\
\hline
Repetition rate & $2~\text{MHz}$\\
\hline
Satellite pass time (ideal) & $\approx 120~\text{s}$\\
\hline
\end{tabular}
\end{center}
\caption{SPOQC mission parameters.}
\label{tab:SPOQC_param}
\end{table}

\begin{widetext}

\begin{figure}[h]
\centering
\includegraphics[scale=0.4]{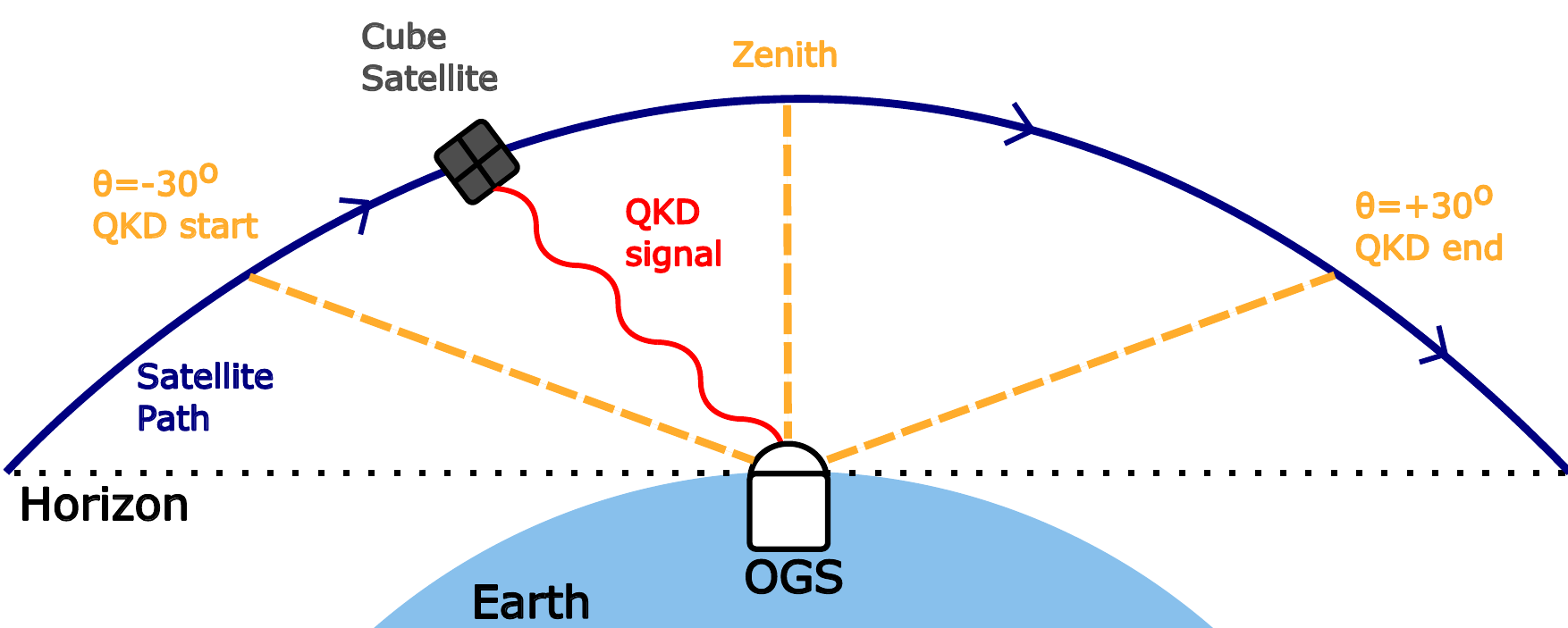}
\caption{Satellite pass over the OGS. Here, the CV-QKD exchange occurs between zenith angles $-30^\circ$ and $+30^\circ$. For this scenario, the satellite will have an altitude of $550~\text{km}$.}
\label{fig:overpass_fig}
\end{figure}

\end{widetext}

\subsection{Satellite Overpass}
\label{sat_pass}
For a satellite in LEO, the channel losses will be highly dynamic as the signal propagation length changes with respect to the zenith angle as the satellite passes over the OGS. The signal propagation path length will be changing with respect to the zenith angle of the satellite, as the satellite passes over the OGS. The propagation path length $z(\theta)$ depends on the zenith angle $\theta$ as \cite{cakaj2011range},
\begin{equation}
z (\theta) =\sqrt{a_{z}^2+2a_{z}r_{e}+r_{e}^2\cos^2(\theta)}-r_{e}\cos(\theta),
\label{eqn:L}
\end{equation}
where $r_e$ is the radius of the earth and $a_{z}$ is the altitude of the satellite. The CV-QKD satellite overpass is shown in Fig. \ref{fig:overpass_fig}. The QKD exchange occurs between zenith angles $-30^\circ$ and $+30^\circ$ for our case. We note here that Eq. (\ref{eqn:L}) is true for a satellite at any orbit and is applicable for any range of zenith angles. Since the effective transmission, or equivalently the effective channel loss, is a function of distance of the satellite from the OGS, it is paramount to characterise the variable channel loss between those angles.

\subsection{Turbulence Profiles and Strengths}
\label{turb_strength}
\begin{figure}
\centering
\includegraphics[width=\linewidth]{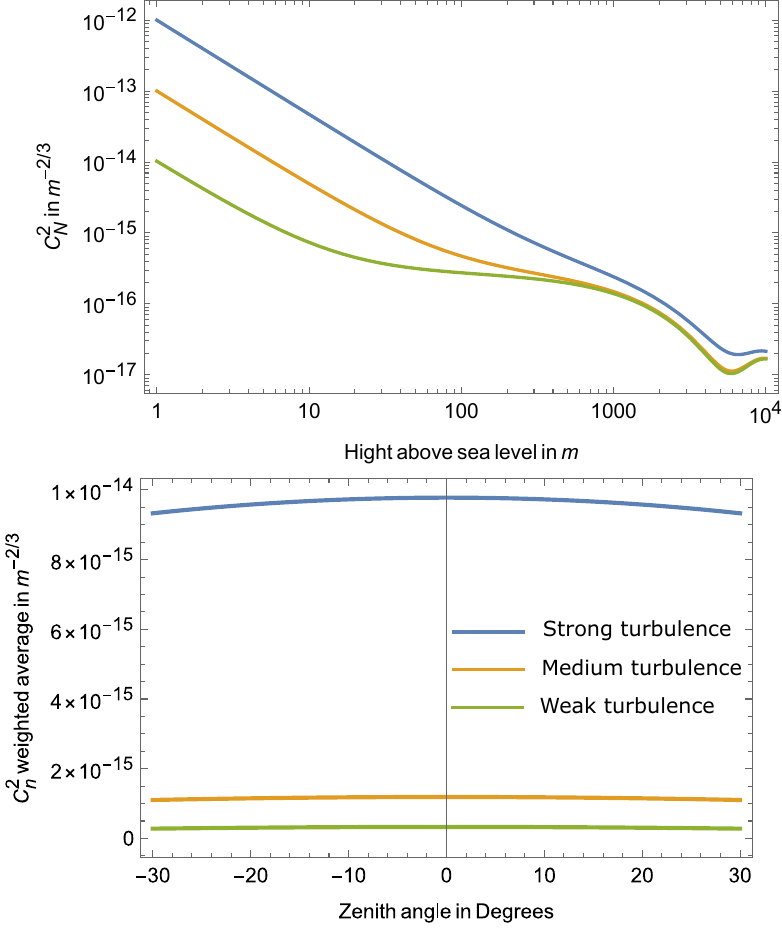}
\caption{$C_N^2$ profiles as a function of altitudes (shown on the top) and the weighted $C_N^2$ profile for $550~\text{km}$ downlink as a function of zenith angles (shown on the bottom). Here the power law parameter for day time $p_{\text{day}}=1.33$ was used, random background turbulence parameter $M=1$, root-mean-square high altitude wind speed $v_w=21~\text{m/s}$, and with the initial turbulence strengths as seen in the parameter table.}
\label{fig:cn2_profiels}
\end{figure}

Turbulence or turbulent flow is the fluid motion characterised by chaotic changes in pressure and flow velocity of the medium. An electromagnetic wave experiences turbulence as it travels in any medium like air and water. Atmospheric turbulence refers to the irregular and chaotic movement of air in the Earth's atmosphere, which is characterised by fluctuations in wind speed and direction, creating a turbulent flow. This turbulence can affect various phenomena, including light propagation, which impacts astronomical observations and optical communication. Thus characterisation of the turbulence is very important factor in both uplink- and downlink-based communication, which is done through the factor of turbulence strength of the medium. The strength of turbulence is conventionally given and measured as a function of the refractive index structure parameter, $C_N^2$ \cite{otoniel2015atmospheric,tyson2022principles} of the medium. In this work, the Hufnagel-Andrew-Phillips model was chosen to calculate the turbulence strength of Earth's atmosphere, as it is more accurate towards ground level and is equivalent to the Hufnagle-Valley model at higher altitudes \cite{andrews2009near}. The turbulence strength according to Hufnagel-Andrew-Phillips model is given by,
\begin{align}
C_N^2(h) = &M\bigg[0.00594\left(\frac{v_w}{27}\right)^2(h\times10^{-5} )^{10}\exp\left(\frac{-h}{1000}\right) \nonumber \\
&+(2.7\times10^{-16})\exp\left({\frac{-h}{1500}}\right) \nonumber \\
&+C_N^2(h_0)\left(\frac{h_0}{h}\right)^p \bigg] ,
\label{eqn:cn2}
\end{align}
where $h$ is the altitude of the respective turbulence layer, $v_w$ is the root-mean-square high altitude wind speed, $h_0$ is the height of the instrument above the ground, $C_N^2(h_0)$ is the average refractive index structure parameter at $h_0$, $M$ is the random background turbulence and $p$ is the power-law parameter depending on the temporal time of the day. From the Hufnagel-Andrew-Phillips model, the turbulence strength for a given turbulence layer can be calculated. The turbulence layer here signifies the respective atmospheric layer which has almost similar turbulence strength. Therefore, the turbulence strength across a turbulence layer is a constant. But in order to find the turbulence profile through a vertical channel, the $C_N^2$ values will be averaged according to the following weighted average \cite{fante1980electromagnetic},
\begin{align}
&I_0^{\text{up}}(\theta)=\int_0^{z(\theta)}d\xi\bigg(1-\frac{\xi}{z(\theta)}\bigg) ^{5/3}C_N^2[h(\xi,\theta)], \nonumber \\
&I_0^{\text{down}}(\theta)=\int_0^{z(\theta)}d\xi\bigg(1-\frac{\xi}{z(\theta)}\bigg)^{5/3}C_N^2[h(z(\theta)-\xi,\theta)],
\label{eqn:I_0}
\end{align}
where $I_0^{\text{up}}$ is the weighted $C_N^2$ average for an uplink (ground-to-satellite) channel and $I_0^{\text{down}}$ is the weighted $C_N^2$ average for a downlink (satellite-to-ground) channel. This value of $C_N^2$ will be used for all the turbulence loss calculations with different turbulence strength, using the ground level turbulence strength $C_N^2(h_0)$ as reference. In this work only a satellite-to-ground link will be considered. The refractive index structure parameters as a function of altitude and weighted averages as a function of zenith angle are shown in Fig. \ref{fig:cn2_profiels}.

\subsection{Atmospheric attenuation}
Atmospheric attenuation corresponds to the weakening or reduction in the intensity of an electromagnetic signal, as it travels through the Earth's atmosphere. This can happen both for uplink and downlink transmission. The attenuation can be due to either absorption or scattering by the gases or particles in the atmosphere. Various techniques can be employed to mitigate the effects of atmospheric attenuation, such as using specific wavelengths or frequencies less susceptible to absorption, employing adaptive techniques, or using signal processing methods to compensate for signal degradation. Here we limit our arguments to only the attenuation due to scattering. Thus, the atmospheric attenuation due to scattering will depend on the extinction factor and is defined as \cite{vasylyev2019satellite},

\begin{equation}
T_{atm}=e^{-\alpha_0g(\theta)},
\end{equation}

where $\alpha_0$ is the extinction factor. This depends on the scattering due to aerosol distribution and Rayleigh scattering, where Rayleigh scattering is wavelength dependent as $1/\lambda^4$. Scattering due to aerosols contributes the same amount as Rayleigh scattering \cite{elterman1968uv, vasylyev2019satellite}. Additionally, 

\begin{equation}
g(\theta)=\int_0^{z(\theta)}e^{-h(\xi,\theta)/\bar{h}}d\xi,
\end{equation}

with $\Bar{h}=6600~\text{m}$ and is a scaling factor. For $1550~\text{nm}$ and a satellite altitude of $550~\text{km}$, the atmospheric attenuation is very minimal at the zenith, only around $1\times10^{-2}~\text{dB}$. 

\subsection{Scintillation}
\label{scin}
One of the main effects of atmospheric turbulence is scintillation. When light propagates through a turbulent media, the beam front will be perturbed and intensity speckles will be created due to the phase differences at different points in the beam. These beam front perturbations can be measured and corrected using adaptive optics. This means scintillation creates intensity fluctuations on the beam front and is conventionally expressed in terms of amplitude variance \cite{tyson2022principles},

\begin{equation}
\sigma_{sci}^2(\theta)=1.23k^{7/6}z(\theta)^{11/6}I_0(\theta),
\label{eqn:scin}
\end{equation}

where $k$ is the wavenumber and $I_0$ is the weighted $C_N^2$ average calculated from Eq. (\ref{eqn:I_0}). For the amplitude scintillation variance $\sigma_{sci}^2\gg1$, the variance due to scintillation corresponds to the high turbulence regime. As seen in the above equation, the scintillation amplitude variance depends on both the propagation distance $z(\theta)$ and the turbulence strength. Therefore, the scintillation amplitude variance increases with distance as well as the turbulence strength. A satellite at low altitude but in high turbulent medium, and another at a high altitude in low turbulent medium could have same variance due to scintillation at the receiver. In satellite-to-ground QKD the propagation distance will always be in the scale of hundreds of kilometre, this means Eq. (\ref{eqn:scin}) will always be greater than one regrades of the initial strength of turbulence at the ground. Physically this means due to the propagation distance the beam will pass through many turbulence layers before reaching the ground. Even if the level of turbulence of each of these layers is low, there will be an compounding effect of each of these layers on the beam. Therefore, the observed scintillation effect on the ground will appear as a high turbulence scintillation due to the long propagation distance. Consequently, satellite-to-ground QKD channels will always be within the high turbulence regime and we shall assume a high turbulence strength for the rest of this work. The relevant high turbulence equations must be used, and thus, the resulting irradiance variance is calculated as \cite{fante1975electromagnetic}, 

\begin{equation} \label{eqn:sigI}
\sigma_{I}^2(\theta) = 1+\frac{0.99}{(\sigma_{sci}^2(\theta))^{2/5}}.
\end{equation}

Using the irradiance variance multiplied by an aperture averaging constant $A$, the variance in loss due to scintillation can be calculated. Here $A$ is defined in the strong turbulence regime as \cite{churnside1991aperture},

\begin{align} \label{eq:Aputure_av}
A &=\frac{\sigma_{I}^2(\theta)^2+1}{2\sigma_{I}^2(\theta)}\bigg(1+0.908\bigg(\frac{D}{2r_0}\bigg)^2\bigg)^{-1} \nonumber \\
&+\frac{\sigma_{I}^2(\theta)-1}{2\sigma_{I}^2(\theta)}\bigg(1+0.162\bigg(\frac{kr_0D}{2z(\theta)}\bigg)^{7/3}\bigg)^{-1},
\end{align}

where $D$ is the receiver aperture diameter and $r_0$ is the Fried parameter and is defined as \cite{tyson2022principles, fante1975electromagnetic,churnside1991aperture},

\begin{equation}
r_{0}=(1.46k^{2}C_{N}^{2})^{-3/5},
\end{equation}

where $k$ is the wavenumber and $C_{N}^{2}$ is the refractive index structure parameter which can be replaced with $I_{0}$ from Eq. (\ref{eqn:I_0}), if instead of horizontal propagation, an uplink or downlink path is considered. For $1550~\text{nm}$, a receiver aperture of $60~\text{cm}$ with $30\%$ obstruction and a satellite altitude of $550~\text{km}$, the irradiance variance due to scintillation at the zenith is around $2\times10^{-2}~\text{dB}$, given high turbulence strength. 

Since the transmission loss variance due to scintillation is highly dependent on receiver aperture size, it will increase with the receiver aperture. This relationship can be seen the aperture averaging equation (Eq. (\ref{eq:Aputure_av})). From this equation, one can see the averaging constant decreases with increasing receiver aperture. When the aperture averaging is then multiplied by the irradiance variance due to scintillation (Eq. (\ref{eqn:sigI})), the overall transmittance decreases and the loss therefore increases. This means with a larger receiver aperture, the total variance of loss will increase.

\subsection{Beam wandering and broadening}
A diffraction-limited electromagnetic beam propagating through a medium follows the inverse square law. During the propagation through a turbulent atmosphere, it additionally experiences beam wandering \& broadening. Beam wandering is due to turbulence eddies larger than the propagating beam width and occurs on a slow time scale, whereas beam broadening is caused by eddies smaller than the beam width and occurs on a fast time scale \cite{gonzalez2023satellite}. While the former causes random displacement in the beam centre, with respect to the centre of the receiver telescope, the latter increases the beam width on the ground. The degree of both beam wandering and broadening depends on the turbulence strength, wavelength and initial beam width, altitude, etc.

The intensity profile of a Gaussian beam is given by \cite{bea1991fundamentals},

\begin{equation}
I(r)=\frac{2P}{\pi \omega (\theta)^2}\exp\bigg(\frac{-2r^2}{\omega(\theta)^2}\bigg),
\label{eqn:gaussian_int}
\end{equation}

where $r$ is the radial distance from the centre of the beam, $P$ is the power of the beam and $\omega(\theta)$ is the propagating beam width and is dependent on the propagation distances as follows \cite{alda2003laser}, 

\begin{equation} \label{eqn:omega}
\omega(\theta)=\omega_{0}\sqrt{1+\bigg(\frac{z(\theta)\lambda}{\pi\omega_{0}^2}\bigg)^2},
\end{equation}

where $\omega_{0}$ is the beam waist, $z(\theta)$ is the propagation distance and $\lambda$ is the wavelength of the light. The propagating beam width $\omega(\theta)$ can be replaced with the short term beam spread $\omega_{st}(\theta)$ in Eq. (\ref{eqn:gaussian_int}) to make the beam width dependent on propagating distance and the turbulence strength \cite{yura1973short} as,

\begin{equation}
\omega_{st}(\theta)=\sqrt{\omega(\theta)^2+2\bigg(\frac{\lambda z(\theta)}{\pi r_0}\bigg)^2(1-\varphi)^2},
\label{eqn:omegast}
 \end{equation}

where $\varphi=0.33\big(\frac{r_0}{\omega_0}\big)^{1/3}$ is a constant. 

Now using these quantities and Eq. (\ref{eqn:gaussian_int}), we calculate the transmittance at a given displacement, with respect to the centre of the receiver telescope, as:

\begin{equation}
T_{\text{dis}} = 2 \int_0^{a_r /2}\int_{d_r-x}^{d_r+x} \frac{2}{\pi \omega_{st} (\theta)^2}\exp(\frac{-2r^2}{\omega_{st}(\theta)^2})drdx,
\label{eqn:T_dis}
\end{equation}

with $d_r = \sqrt{d^2+(a_r/2-x)^2}$, $r$ is the radial distance from the centre of the beam, $d$ is the displacement from the centre of the beam and $\omega_{st}(\theta)$ is given in Eq. (\ref{eqn:omegast}) and $a_r$ is the receiver aperture. Since $T_{\text{dis}}$ is dependent on the propagating beam width, it calculates the total loss due to beam broadening and displacement. The derivation for this equation is shown in Appendix \ref{appendix}.

To determine the deviation $d$ of the beam from the centre of the receiver aperture, beam wandering effects must be investigated. Loss due to beam wandering can be separated into two different parts - beam wandering due to turbulence and pointing error of the satellite. The turbulence component and pointing error is observed to vary with a probability of a Weibull distribution. The displacement variance due to turbulence is by the following equation \cite{fante1975electromagnetic, yura1973short},

\begin{equation}
\sigma_{TB}^2=\frac{0.1337\lambda^2z(\theta)^2}{\omega_0^{1/3}r_0^{5/3}},
\label{eqn:sig_TB}
\end{equation}

where $\lambda$ is the wavelength, $z(\theta)$ is the propagation path length, $\omega_0$ is the beam waist and $r_0$ is the Fried parameter. In addition to this variance in Eq. (\ref{eqn:sig_TB}), the deviation from the centre $d$ will also depend on the pointing error. This error is derived from the tracking error at the satellite (known as pointing error) as,

\begin{equation}
d_{\text{Tr}}=a_z\tan{\bigg(\frac{t_{\text{err}}}{2}\bigg)},
\label{eqn:d_err}
\end{equation}

where $a_z$ is the altitude of the satellite and $t_{\text{err}}$ is the pointing error. Now, Eq. (\ref{eqn:sig_TB}) and Eq. (\ref{eqn:d_err}) can be added to obtain the total deviation of the beam from the centre of the receiver aperture $d$ as,

\begin{equation}
d=\sigma_{TB}+d_{\text{Tr}}.
\label{eqn:d}
\end{equation}

For $1550~\text{nm}$, a receiver aperture of $60~\text{cm}$ with $30\%$ obstruction, a transmitter aperture of $8~\text{cm}$ and with the satellite at an altitude of $550~\text{km}$, the loss at the zenith from beam boarding \& wandering due to turbulence is $2\times10^{-4}~\text{dB}$ and $1\times10^{-2}~\text{dB}$ respectively. This is given for a high turbulence strength. The geometrical loss due to beam diffraction, given the same parameters used above, is around $29.1~\text{dB}$ at the zenith. 

\subsection{Tracking}
In order to establish and maintain an optical link over the dynamic trajectory, the satellite and OGS rely on uplink and downlink tracking beacon lasers at two different wavelengths. One can determine the angular offset that makes the OGS or satellite go out of the beam area of the respective beacon lasers. This quantifies the tracking accuracy required for a given beacon wavelength, satellite altitude and beacon laser telescope aperture. The error in tracking can be deconstructed into two parts: (a) error due to the satellite beacon being offset with respect to the OGS, referred to as \textit{pointing error}; (b) error due to the OGS being offset with respect to the satellite- referred to as \textit{coarse tracking error}. For example, a satellite at an altitude of $550~\text{km}$ with the downlink beacon laser of $685~\text{nm}$m using $8~\text{cm}$ (or $2~\text{cm}$) aperture optics has the pointing error of around $1.1~\mu\text{rad}$ (for either aperture size). For an OGS with $60~\text{cm}$ (or $2~\text{cm}$) aperture and $850~\text{nm}$ uplink beacon laser, the tracking error is $1.2~\mu\text{rad}$ (or $1.5~\mu\text{rad}$).
 
The error in tracking by the OGS will affect how well the received signal will couple to the detector, referred to as \textit{fine tracking error}. Since the fine tracking requirements are highly dependent on the angle of the field of view (AFOV), they will depend on the focal length of the telescope used as well as the size of the quantum detector/sensor. For a given telescope of $60~\text{cm}$ aperture with $6469~\text{mm}$ focal length, the minimum amount of tracking required to couple the light to single mode fibre (SM fibre) is $0.8~\mu\text{rad}$, for multi mode fibre (MM fibre) it is $6~\mu\text{rad}$. For a large area free space CV receiver (CVR) with a detection size of $3~\text{mm}$, the accuracy of the required tracking is $0.2~m\text{rad}$. The use of a large area free space is advantageous, as the tracking requirements are much reduced compared to SM and MM fibre \cite{kumar2021increasing}. The tracking requirements for a large free space detector is much larger than the minimum pointing accuracy's of $1.5~\mu\text{rad}$ required from the tracking beacons.

To mitigate the losses due tracking there are two different main methods of implementing tracking, open- and close-loop tracking \cite{khairallah2021ephemeris}. In both of these tracking methods, beacon lasers, tracking cameras, and fine steering mirrors (FSM) are used. For open-loop tracking the tracking algorithm is predetermined and a downlink beacon is not strictly necessary. For close-loop tracking, the offset is measured and corrected using a downlink beacon and FSM. Closed-loop tracking will have more overhead, but will have a much greater accuracy. 

\begin{figure}
\centering
\includegraphics[width=\linewidth]{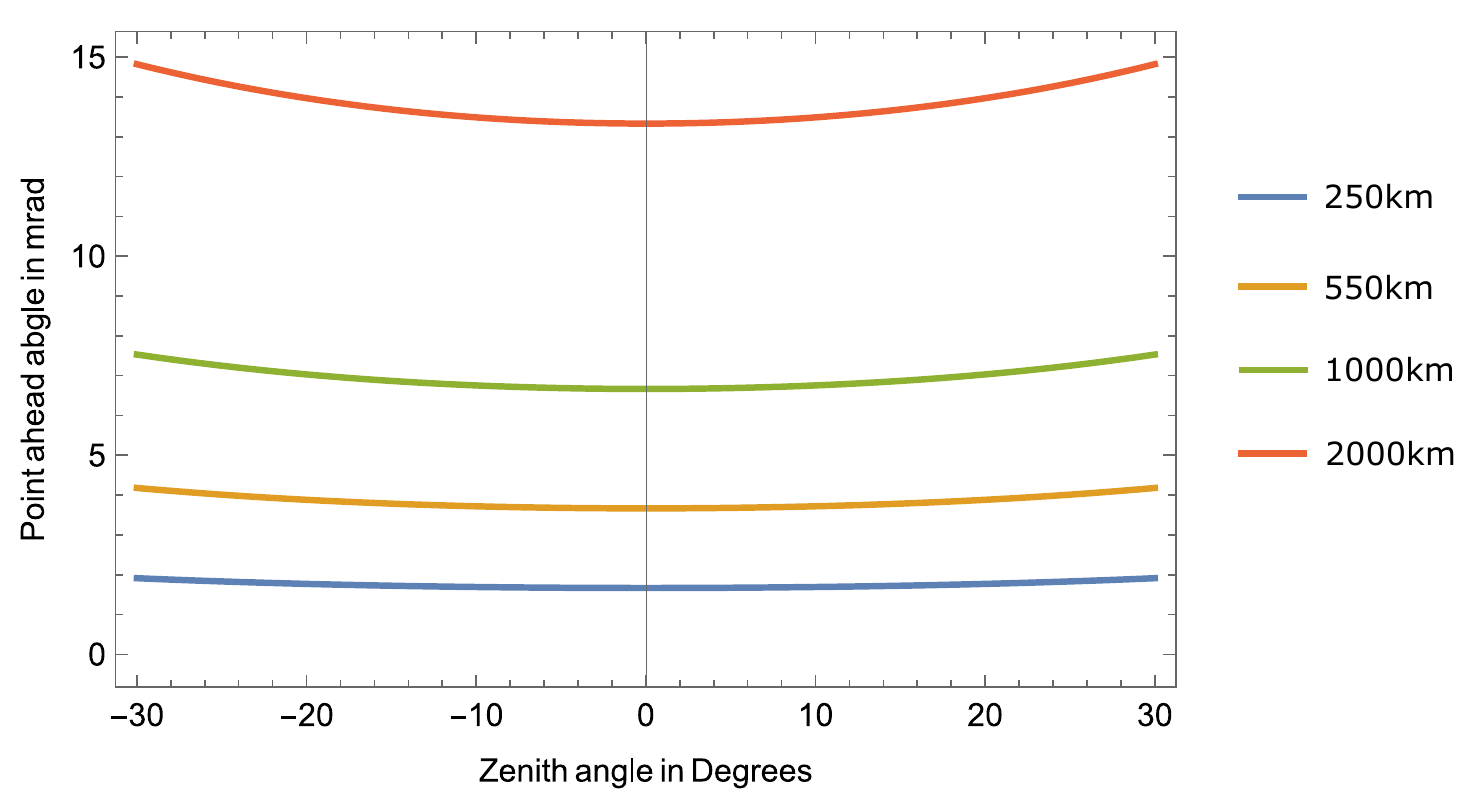}
\caption{Point ahead angle for a given satellite altitude. The PAA changes with increasing altitudes of the satellite, ranging from $250~\text{km}$ to $2000~\text{km}$. This is a typical range for LEO satellites.}
\label{fig:PAA}
\end{figure}

In order to improve tracking further, the point ahead angle (PAA) can be used. Due to the relative motion between satellite and OGS, and the time taken by the light to travel between them, the beacon lasers are sent with an angular offset known as PAA \cite{lazzaro2022evaluation},

\begin{equation}
\theta _{PAA}=2 \frac{v_t}{c},
\label{eqn:PAA}
\end{equation}

where $c$ is the speed of light and $v_t$ is the tangential velocity of the satellite. Here the PAA at a given zenith angle is shown in the Fig. \ref{fig:PAA} for varying altitudes of the satellite. This shows that the point ahead angle increases with increasing altitude of the satellite. As seen in this figure, even at low altitudes, the PAA is in the order of $m\text{rad}$, much larger than the minimum pointing accuracy required for the satellite and OGS to continue tracking. It is therefore necessary to employ the use of PAA whilst tracking. The PAA alone is not enough to compensate for pointing error and it cannot be implemented without looped tracking. If the PAA is used in conjunction with closed-loop tracking, then it is possible to reduce the tracking error to around $3~\mu\text{rad}$ \cite{han2018point}.

For $1550~\text{nm}$, a receiver aperture of $60~\text{cm}$ with $30\%$ obstruction, a transmitter aperture for the quantum signal of $8~\text{cm}$ and with the satellite at an altitude of $550~\text{km}$, the loss at the zenith due to pointing is $2\times10^{-1}~\text{dB}$, if a pointing error of $4~\mu\text{rad}$ at the satellite is considered. 

\subsection{Total loss in the satellite-to-ground channel under clear sky conditions}
\label{tot_loss}
The total loss is the combined loss from various sources discussed in this section and the loss within Bob's system, such as optical losses and loss due to detector coupling efficiency. Detector coupling efficiency is aggravated by atmospheric turbulence, and this can be mitigated using large area detectors or with the use of adaptive optics. If adaptive optics is not implemented, the loss introduced by coupling a beam through a turbulent free space channel to fibre is around $13~\text{dB}$, while coupling to a large area free space detector does not induce any additional loss \cite{kumar2021increasing}. In this section the total loss is limited to the loss only due to the channel under clear-sky assumptions.

An additional factor to consider when analysing loss is the ellipticity of the beam at the OGS. This ellipticity arises due to the angle between the satellite and the OGS. Specifically, the path length for the beam's edges nearest and furthest from the horizon will differ, causing the beam to appear elliptical instead of circular at the OGS. The resulting ellipticity impacts loss calculations as the beam width will vary across its edges. According to Eq.(\ref{eqn:T_dis}), the loss is dependent on the beam width at the OGS, and as shown in Eq.(\ref{eqn:omega}), the beam width at the OGS is highly influenced by the propagation distance. The edge of the beam closer to the horizon experiences a shorter propagation distance compared to the edge further from the horizon. This effect becomes more pronounced as the satellite moves away from the zenith, where there is no difference in propagation distance between the edges. Eq.(\ref{eqn:omega}) can be applied to calculate the corresponding difference in beam width with the help of trigonometry. For example, at a zenith angle of $+30^{\circ}$ with a diffraction-limited beam at $1550~\text{nm}$, a satellite altitude of $550~\text{km}$, and a satellite telescope aperture of $8~\text{cm}$, the difference in propagation distance is approximately $9$ meters. This results in a beam width difference of about $0.2~\text{mm}$ at the OGS, compared to an overall beam width of approximately $15$ meters. Given these parameters, the effect of beam ellipticity on loss calculations is negligible, allowing the assumption of a circular beam at the OGS for simplification.

\begin{figure}
\centering
\includegraphics[width=1\linewidth]{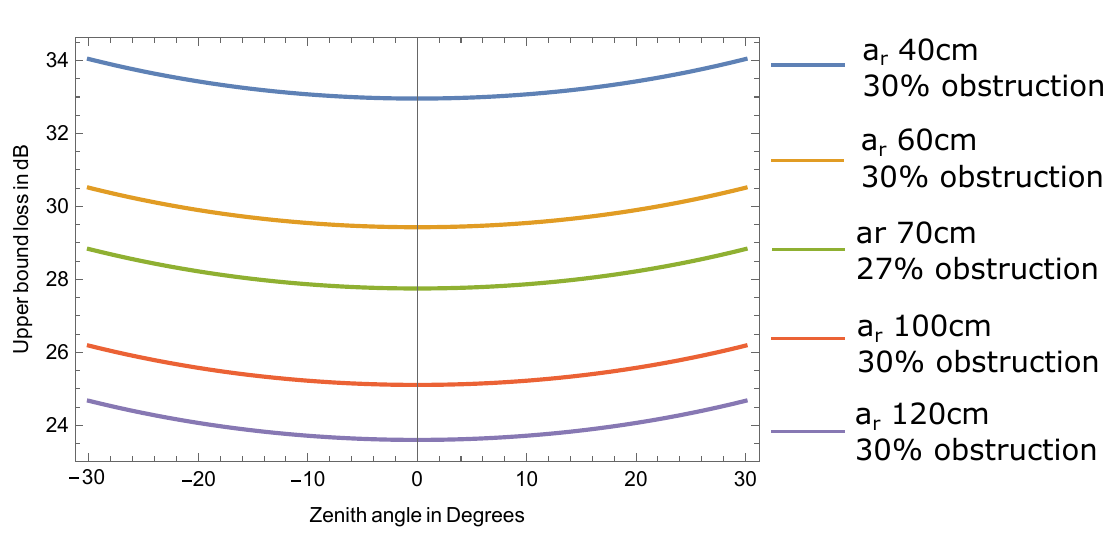}
\caption{Total upper bound channel loss with differing receiver aperture sizes. This was done for a strong turbulence regime, with a wavelength of $1550~\text{nm}$, a transmitter aperture of $8~\text{cm}$, a pointing error at the satellite of $4~\mu\text{rad}$, and with the satellite at an altitude of $550~\text{km}$.}
\label{fig:changing_ar}
\end{figure}

The total channel loss for various different receiver telescope apertures are shown in Fig. \ref{fig:changing_ar}, here the losses are radially symmetric around the OGS site. The upper bound on loss was considered for each of the receiver aperture values given a high turbulence regime. The total loss for the CV-QKD exchange fluctuates around $1~\text{dB}$, for a receiver aperture of $60~\text{cm}$, where it is at a minimum at the zenith and increases with $\theta$.

The majority of losses stem from diffraction, and these losses can be reduced with larger receiver apertures as shown in Eq. (\ref{eqn:T_dis}). Although the loss variance due to scintillation would increase with increasing receiver aperture size. The loss due to scintillation is still comparatively smaller than the loss due to diffraction.

\begin{figure}
\centering
\includegraphics[width=1\linewidth]{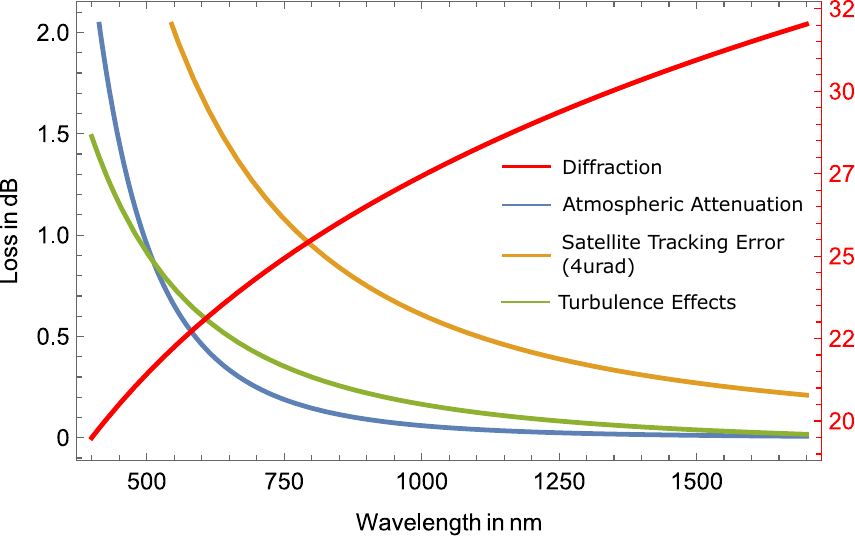}
\caption{The losses as a dependency of wavelength at the zenith is plotted. Here the atmospheric absorption lines have not been taken into consideration. The losses are plotted for a receiver aperture of $60~\text{cm}$ with $30\%$ obstruction, a transmitter aperture of $8~\text{cm}$, a pointing error at the satellite of $4~\mu\text{rad}$, and with the satellite at an altitude of $550~\text{km}$. This is the loss for a Gaussian modulated CV (GMCV) QKD protocol. For DV-QKD, there will be additional losses due to phase error \cite{lo1999unconditional, shor2000simple,koashi2009simple} that GMCV does not need to consider as the transmitted local oscillator will experience the same phase drift as the quantum signal \cite{qi2015generating, jouguet2013experimental,daniel2015self}.}
\label{fig:oss_vs_lambd}
\end{figure}

As seen in all the previous parts of this section, all the losses have a wavelength dependency. This dependency is illustrated in Fig. \ref{fig:oss_vs_lambd}, and the figure shows all the contributions to the total loss individually. Note that the atmospheric absorption lines have been ignored. Here, the total loss will increase with increasing wavelength. This is due to diffraction, which has the largest contribution to the total loss. However, the variance in the loss decreases with an increase in wavelength. The losses that have a variance at a fixed zenith angel are those due to tracking and turbulence. Here, beam wandering due to turbulence and pointing error varies at a scale of $~10~\text{ms}-100~\text{ms}$ \cite{pirandola2021satellite, bourgoin2013comprehensive} and scintillation at a scale of milliseconds \cite{dravins1997atmospheric,osborn2015atmospheric}. The loss due to diffraction and atmospheric attenuation on the other hand will be static at a given zenith angle and will only vary with respect to the zenith angle. Longer wavelengths have the advantage of low loss variance. The variance in loss, such as from tracking error and turbulence, decrease with increasing wavelength, Eq. (\ref{eqn:sigI}), (\ref{eqn:omegast}), (\ref{eqn:T_dis}) and (\ref{eqn:d}). It also implies, in an uplink QKD exchange, that larger wavelengths are more favourable, as turbulence effects on the quantum signal will significantly decrease.

\subsection{Total loss in the satellite-to-ground channel under other weather conditions}
\label{tot_loss_weather}

In the previous subsection, Sec. \ref{tot_loss}, the total loss for various parameters was analysed under clear sky conditions. During realistic satellite-to-ground QKD scenario clear sky conditions cannot be guaranteed for the key exchange, analysis of other weather effects and conditions is vital. In this subsection the adverse effect of weather will be analysed and the total loss for satellite-to-ground CV-QKD will be shown for various weather conditions. We consider the weather condition to be static during the satellite pass, such that the weather condition will not change throughout the $120~\text{s}$ satellite pass. Although we considered the possibility for there to be partial obstructions during pass, i.e. given partial cloud cover.

When considering attenuation from weather conditions such as cloud cover or fog, it is vital to classify the type of cloud/fog in order to determine the water vapour distribution as well as the vertical extent \cite{lyras2016cloud, moll2007wavelength}. Extensive studies have been undertaken in order to calculate the losses due to clouds and fog for optical wavelengths, here we have utilised the results from Moll et. al. \cite{moll2007wavelength}. We will consider best case cloud cover scenario of low clouds (Stratus and Stratocumulus) with their minimal vertical extent of $200~\text{m}$. These clouds not only have the lowest attenuation per kilometre, but also the shortest vertical extent. However, even in this lowest loss case, CV-QKD key generation is not possible. We will also consider different fog conditions. In most of the cases secure key generation is possible only given moderate fog.

\begin{widetext}

\begin{table}[h]
\begin{center}
\begin{tabular}{||c c c c c c||} 
\hline
~Receiver telescope aperture ~&~ Clear sky ~&~ Moderate fog ~&~ Heavy fog ~&~ Stratocumulus clouds ~&~ Stratus cloud ~\\ [0.5ex] 
\hline\hline
$40~\text{cm}$ & $33~\text{dB}$ & $37.3~\text{dB}$ & $45.8~\text{dB}$ & $66~\text{dB}$ & $83.3~\text{dB}$\\ 
\hline
$60~\text{cm}$ & $29.4~\text{dB}$ & $33.7~\text{dB}$ & $42.2~\text{dB}$ & $62.4~\text{dB}$ & $79.7~\text{dB}$\\ 
\hline
$70~\text{cm}$ & $27.8~\text{dB}$ & $32.1~\text{dB}$ & $40.6~\text{dB}$ & $60.8~\text{dB}$ & $78.1~\text{dB}$\\ 
\hline
$100~\text{cm}$ & $25.1~\text{dB}$ & $29.4~\text{dB}$ & $37.9~\text{dB}$ & $58.1~\text{dB}$ & $75.4~\text{dB}$\\ 
\hline
$120~\text{cm}$ & $23.6~\text{dB}$ & $27.9~\text{dB}$ & $36.4~\text{dB}$ & $56.6~\text{dB}$ & $73.9~\text{dB}$\\ 
\hline
\end{tabular}
\caption{Total loss for $1550~\text{nm}$ at Zenith for various receiver telescope apertures and weather conditions. Here, the considered clouds have a vertical extent of $200~\text{m}$, the minimal vertical extent for these type of clouds.}
\label{tab:weather}
\end{center}
\end{table}

\end{widetext}

As seen in Tab. \ref{tab:weather}, the total loss of the channel for a signal of $1550~\text{nm}$ when non-clear sky conditions are considered is very high.
Given moderate fog, the total channel loss is just below $35~\text{dB}$ for all but the smallest receiver aperture size.

Considering other wavelengths for the CV-QKD signal, it may be possible to overcome weather constraints. Shorter wavelengths generally will have less overall loss as seen in the previous section (Sec. \ref{tot_loss}), although the attenuation from water vapour in clouds and fog will increase. Following on from the results in Moll et. al. \cite{moll2007wavelength}, one can approximate the losses to be expected for SPOQC given other signal wavelengths. The expected loss of the channel at the zenith with a $60~\text{cm}$ receiver telescope,at various wavelengths, can be seen in Tab \ref{tab:weather_lambda}. Again the increased loss due to adverse weather conditions is very high. Realistically, the additional losses due to cloud cover makes a QKD key exchange unfeasible. It is be possible for CV-QKD to occur during moderate fog conditions, as given in Tab. \ref{tab:weather_lambda}.

\begin{widetext}

\begin{table}[h]
\begin{center}
\begin{tabular}{||c c c c c c||} 
\hline
~Signal wavelength ~&~ Clear sky ~&~ Moderate fog ~&~ Heavy fog ~&~ Stratocumulus clouds ~&~ Stratus cloud ~\\ [0.5ex] 
\hline\hline
$630~\text{nm}$ & $23.8~\text{dB}$ & $32.5~\text{dB}$ & $38.8~\text{dB}$ & $60.8~\text{dB}$ & $84.8~\text{dB}$\\ 
\hline
$810~\text{nm}$ & $24.9~\text{dB}$ & $33.2~\text{dB}$ & $39.7~\text{dB}$ & $59.9~\text{dB}$ & $84.9~\text{dB}$\\ 
\hline
$1064~\text{nm}$ & $26.6~\text{dB}$ & $30.5~\text{dB}$ & $39.2~\text{dB}$ & $58.6~\text{dB}$ & $75.6~\text{dB}$\\ 
\hline
$1550~\text{nm}$ & $29.4~\text{dB}$ & $33.7~\text{dB}$ & $42.2~\text{dB}$ & $62.4~\text{dB}$ & $79.7~\text{dB}$\\ 
\hline
\end{tabular}
\caption{Total loss for $60~\text{cm}$ telescope at Zenith for various signal wavelengths and weather conditions. Here, the considered clouds have a vertical extent of $200~\text{m}$, the minimal vertical extent for these type of clouds.}
\label{tab:weather_lambda}
\end{center}
\end{table}

\end{widetext}

During partial cloud coverage, key generation is affected by the reduction in usable data for parameter estimation; however its impact is minimal if the CV-QKD system clock rate is comparatively higher. Another way to mitigate the influence of weather is by incorporating OGS site diversity \cite{anipeddi2025optical} in satellite scheduling.

\section{Secret Key Rate} \label{key_rate}

\begin{figure}
\centering
\includegraphics[width=1\linewidth]{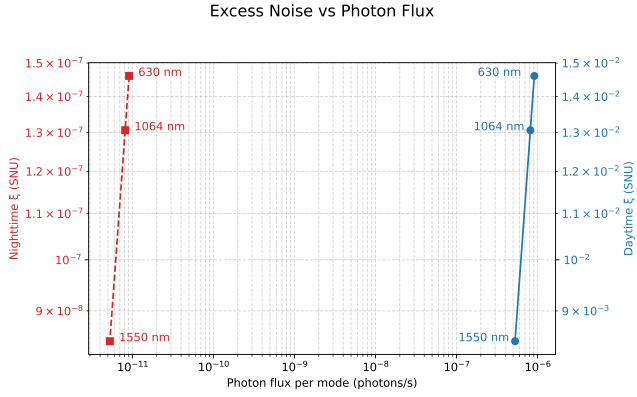}
\caption{Excess noise in CV-QKD ($\xi_{\text{background}}$) as a function of photon flux per homodyne mode is plotted here. The free space homodyne detector of aperture 3 mm is considered, along with laser linewidth of 1$\mu$m and detection efficiency of 0.8. The full calculation of photon flux and resulting excess noise can be seen in Appendix \ref{app:flux}.}
\label{fig:noisevsflux}
\end{figure}

\begin{figure}
\centering
\begin{subfigure}{0.5\textwidth}
\includegraphics[width=0.95\linewidth]{Keyrate_vs_lambda1.jpeg}
\caption{}
\label{fig:SKR2a} 
\end{subfigure}
\medskip
\begin{subfigure}{0.5\textwidth}
\includegraphics[width=0.95\linewidth]{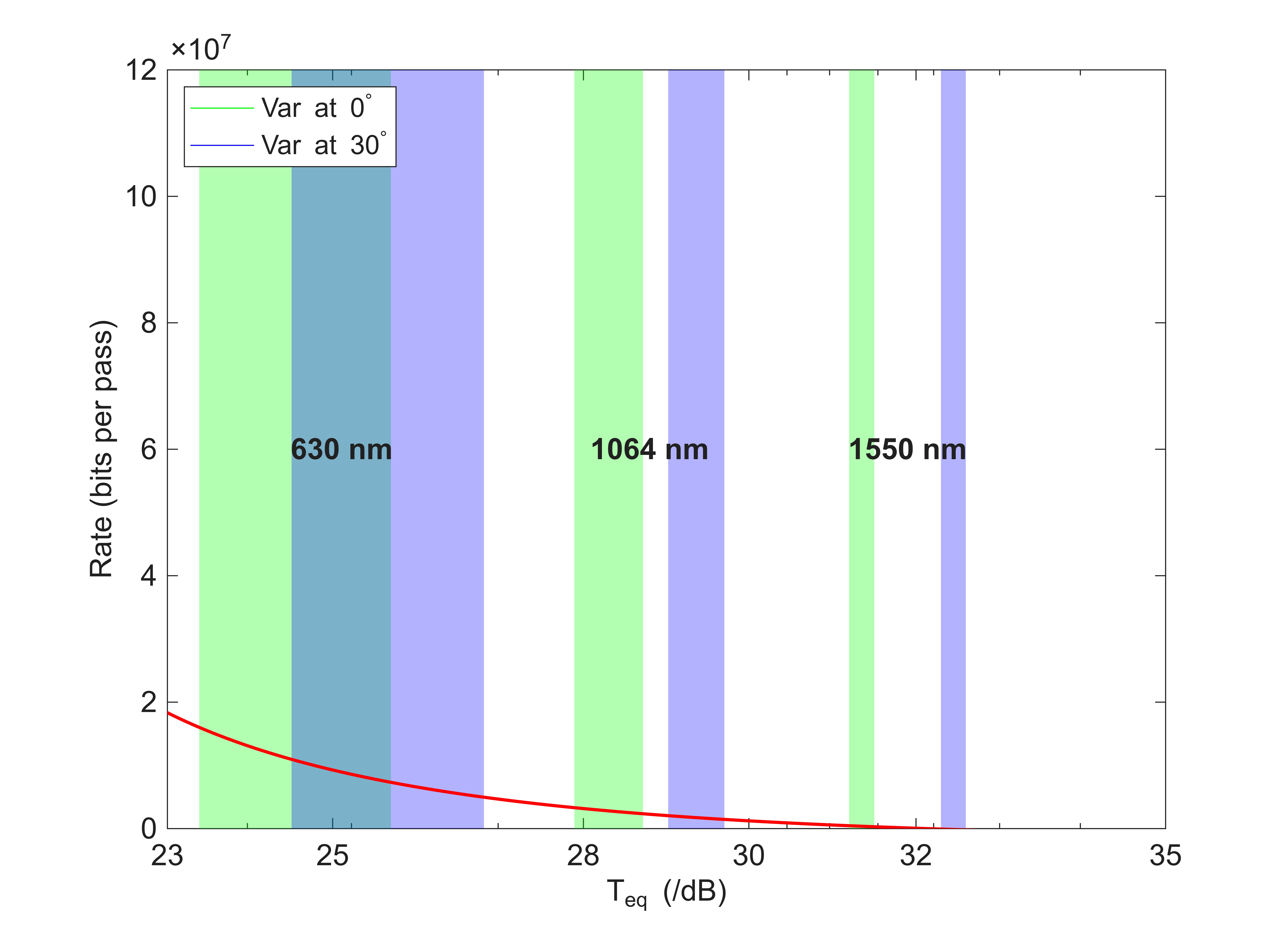}
\caption{}
\label{fig:SKR2b}
\end{subfigure} 
\caption{The achievable secret key rates per pass for the given parameter values as a function of channel loss are plotted above. The solid lines indicate the key rate with restricted Eve assumption, with parameters given in Table \ref{tab:keyrate}, where Fig. \ref{fig:SKR2a} (Fig. \ref{fig:SKR2b}) indicating possible key rates during nighttime (daytime). The green (blue) shaded region represents the variance of loss at zenith ($\pm30^{\circ}$).}
\label{fig:SKR2}
\end{figure}

For the asymptotic regime in CV-QKD, considering reverse reconciliation, the secret key rate is given by \cite{laudenbach2018continuous},

\begin{equation}
K=\beta I_{AB} - \chi_{BE},
\end{equation}

where $\beta$ is the reconciliation efficiency, $I_{\text{AB}}$ is the mutual information between Alice and Bob and $\chi_{\text{BE}}$ is the Holevo information between Bob and Eve. From the channel loss, the secret key rate can now be estimated for a downlink based CV-QKD protocol. The maximum loss for this estimation was around $33~\text{dB}$. In order to calculate the secret key rate for this scenario, more realistic restrictions on Eve were considered. A more realistic secret key rate estimation for space-based QKD was proposed in \cite{ghalaii2023satellite}, line of sight key exchange in the presence of a bypass channel (which Eve has no access to). Here, the assumption is that Eve cannot place herself arbitrarily close to Alice at the satellite or Bob at the OGS. Therefore, she will also have a lossy channel, with a bypass channel that she has no access to. This means not all of the information loss observed by Alice and Bob is leaked to Eve. The consequence of this is that Alice and Bob may increase the modulation variance without leaking too much information to Eve, and thereby increasing their signal to noise ratio. This will in turn increase their secret key rate. Since the consideration only imposes restrictions on Eve, the mutual information between Alice and Bob is defined in the conventional way and for a homodyne detection scheme is given by \cite{laudenbach2018continuous},

\begin{equation}
I_{AB}= \frac{1}{2}\log_{2}\frac{V+\chi_{\text{tot}}}{1+\chi_{\text{tot}}},
\label{eq:iab}
\end{equation}

where $V$ is the modulation variance $\chi_{\text{tot}} = \chi_{\text{line}} +\chi_{\text{hom}}/\eta_{\text{ch}}$ is the total noise at Alice, $\chi_{\text{line}}=(1-\eta_{\text{ch}})/\eta_{\text{ch}}+\xi_{\text{tot}}$ is the noise in the channel, $\chi_{\text{hom}}=(1-\eta_{d})/\eta_{d}+v_{\text{el}}/\eta_{d}$ is the noise term in the homodyne detector with $\eta_{\text{ch}}$ being the transmissivity of the channel, $\xi_{\text{tot}}$ being the excess noise at Alice, $\eta_{d}$ being the detector efficiency and $v_{\text{el}}$ denoting the electronic noise in the receiver. In order to establish a secret key, the mutual information between Alice and Bob must be calculated. As seen in Eq. (\ref{eq:iab}) this depends on the transmissivity (calculated in Sec. \ref{channel_loss}), total excess noise and detector settings.
The expression for total excess noise in a free space CV-QKD system is shown below \cite{laudenbach2018continuous},

\begin{equation}
\xi_{\text{tot}} = \xi_{\text{doppler}} + \xi_{\text{turb}} + \xi_{\text{para}} + \xi_{\text{background}},
\end{equation}

where $\xi_{\text{doppler}}$ is the excess noise due to Doppler effect and arises from frequency mismatch between LO and signal when LLO implementation is considered and can be seen has added phase error \cite{schlake2025pulse, soh2015self}. $\xi_{\text{turb}}$ is the excess noise from phase error due to turbulence effects when LLO implementation is considered. If LLO implementation is used, the Doppler effect and wavefront errors will have to be actively compensated or compensated post measurement with Digital Signal Processing (DSP) at the ground station, this adds a lot of complexity to the OGS set-up. All correction methods will still incur some amount of error/excess noise and one will not be able to mitigate these effects completely. If instead a TLO is used, these sources of errors/excess noise can be avoided completely. $\xi_{\text{para}}$ is the excess noise due to poor parameter estimation. This source of excess noise can be much larger than conventionally assumed due to the dynamic nature of the satellite-to-ground channel. The transmissivity of the channel will vary with zenith angle and additionally have a variance at any given zenith angle (see Fig \ref{fig:changing_ar} \& \ref{fig:oss_vs_lambd}). Larger variance in transmissivity leads to higher parameter estimation error. $\xi_{\text{background}}$ is the excess noise due to background noise photons and depends on the chosen wavelength, time of day and field of view (FOV) of the receiver. The excess noise from background radiation could be a major hurdle for daytime QKD operations \cite{chu2021feasibility, li2023free, cai2024free, zhan2025long}. Assuming the detector efficiency of 0.8, free space detector of aperture 3 mm, the laser linewidth of 1$\mu$m, and an ambient temperature of 300 K, we find that the number of photons in the detection window of 2 ns is of the order $\approx10^{-1}$ ($\approx10^{-6}$) during daytime (nighttime) for 1550 nm. However, we find that the photon flux per homodyne mode is of the order $\approx 10^{-7}$ ($\approx 10^{-12}$), see Appendix \ref{app:flux} for calculation of photon flux and excess noise. This is because the LO acts as a natural filter to select the part of received light that is coherent with LO alone, in a homodyne setup. This is plotted in Fig. \ref{fig:noisevsflux}, where we observe that the background noise at nighttime is negligible compared to other contributions for excess noise. However, we count it is an important factor for daytime operation of CV-QKD, which is also shown in Fig. \ref{fig:SKR2}.

The detector characteristics need to be addressed as well. When considering a free space based CV-QKD scheme, the issue of mode mismatch between the LO and signal arises at the detector. This will cause additional losses to the system. The mode mismatch can be spacial or temporal, and for space based CV-QKD it arises from Doppler and turbulence effects. When considering the Doppler effect, the difference in time of arrival between LO and signal causes temporal mode mismatch at the detector \cite{schlake2025pulse}. Spacial mode mismatch arises form turbulence effects, specifically scintillation. As seen in Sec. \ref{scin}, scintillation will cause wavefront aberrations. These wavefront errors will then cause mode mismatch between the LO and signal. Adaptive optics can be used to correct the wavefront aberrations, although only to the Strehl ratio \cite{tyson2022principles}. In classical free-space optical communication, TLO combined with coherent homodyne detection is a practice to mitigate scintillation effects \cite{al2020survey}. The important quantities of a detector itself are the detection efficiency $\eta_d$, electronic noise $v_{el}$ and detector area. The detector area may have particular importance, if adaptive optical techniques are not used \cite{peyronel2016luminescent}. For example, a large area free space detector of $3~\text{mm}$ diameter saves $13.5~\text{dB}$ loss in terms of coupling loss compared to a single mode fibre \cite{kumar2021increasing}. Coupling to the fibre based detectors can be improved using adaptive optics, although this increases the complexity of the receiver system at Bob. Adaptive optics will also only correct the incoming beam to the limit of the Strehl ratio and coupling to fibre will always induce loss \cite{tyson2022principles}. Although large area detectors improves the coupling ratio, the operating bandwidth of the CV-QKD system will be reduced to a few 10's of MHz \cite{kumar2021increasing}. However, it is within the signal generation bandwidth of the CV-QKD payload in the SPOQC mission. Detector size will also change the FOV of the detector, and thereby the amount of excess noise due to background noise photons.

For the SPOQC CV-QKD payload, the choice of TLO implementation not only reduces protocol complexity but also significantly avoids many sources of excess noise. With a TLO, no excess noise is introduced by Doppler shifts, or turbulence-induced phase errors, since the LO and signal co-propagate through the same channel. In this configuration, both LO and signal undergo the same frequency shifts from Doppler effects, eliminating Doppler-induced noise. Similarly, turbulence affects both identically, so for the interference at the shot noise-limited homodyne detector, both remain coherent with respect to each other. In other words, signal and LO remain in phase. For LLO the OGS set-up complexity is greatly increased, and even though the sources of excess noise due to turbulence and Doppler effect can be mitigated, they cannot be completely eliminated. The TLO set-up has much lower complexity and was therefore chosen for the SPOQC mission, as this is the first proof of concept mission for satellite-to-ground CV-QKD. For the SPOQC mission, the laser has LO launch capability 100's of Watts to one killo Watt of peak power, which is adequate to make the homodyne detector at the OGS shot noise limited over a 30 dB - 40 dB channel. In order to mitigate leakage, the LO is time and polarisation multiplexed with respect to the signal. The main disadvantage is that TLO does have the concerns for security loopholes,  although loopholes also exist for LLO as well \cite{ren2019reference}.

Along with the mutual information between Alice the Bob, another vital quantity necessary for calculating the secret key is the Holovo information. For the restricted Eve in presence of a bypass channel scenario, Eve's channel is restricted, the Holevo information is now dependent on the channel loss between Alice and Eve, $\eta_{AE}$. This means the covariance matrix (CM) established between all the parties differs from the conventional CM by its elements relating to Eve, and thereby, the Helovo information will also differ accordingly. The CM elements relating to Eve will now depend on $\eta_{AE}$. The CM and Holevo bound for the bypass channel restricted Eve are given in the Appendix \ref{apend:bypass}. Following the key rate assumptions in Ghalaii et. al., \cite{ghalaii2023satellite}, the bypass channel parameters are defined as follows; $\eta_{AE}$ is the transmissivity of the channel between Alice and Eve which Eve has access to, $\eta_{S}$ is the transmissivity of the bypass channel which is considered to be a pure loss channel which Eve has no access to, and $\eta_{T}$ is the coupling beam splitter transmissivity of the telescope in Bob's OGS. These parameters can now be optimised in order to generate a positive secret key for a given channel loss.

Under ideal conditions, such as excess noise $\xi_{\text{tot}}=0.001$, reconciliation efficiency $\beta=1$, detector efficiency of $\eta_{d}=1$, electronic noise $v_{\text{el}}=0$, optimal modulation variance $V_{\text{opt}}=300~\text{SNU}$ and telescope coupling transmissivity $\eta_{T}=0.1$. A positive key rate can only be established if the loss between Alice and Eve channel is restricted to $\eta_{AE}=0.05$. This is the minimum amount of restriction needed of Eve in order to generate a positive secret key for the channel loss of a satellite-to-ground channel. Physically, this means Eve will need to be at least around $210~\text{km}$ away from Alice, so that Alice \& Bob can establish a positive secret key. The distance Eve is from Alice was calculated using Eq. (\ref{eqn:T_dis}), and for simplification, only loss due to diffraction was considered. The loss $T_{\text{dis}}$ was equated with $\eta_{AE}$ and the corresponding distance was calculated. This means geometrically Eve is directly between Alice and Bob ($\eta_{S}=0$). This is assuming Eve is in the centre of the quantum signal and has a telescope size comparable to Bob's, i.e., there will be no bypass channel and all of the signal Bob receives will have gone through Eve first. If on the other hand a bypass channel is assumed ($\eta_{S}>0$), then Eve is not in the centre of the beam between Alice and Bob. A positive secret key can not be established under this assumption, but measures can be taken to avoid this as mentioned in the ref \cite{ghalaii2023satellite}.  In our analysis, we have considered $\eta_{AE}=0.05$ which is corresponding to $210~\text{km}$ away from Alice. Given the altitude of Alice's satellite of $550~\text{km}$, Eve will be at an maximum altitude of $340~\text{km}$ above sea level. At this altitude Eve can reasonably be in a Very Low  Earth Orbit (VLEO) or Eve can be on a High Altitude Platform (HAP), aircraft or drone between Alice and Bob. One can reasonably assume Eve utilising such systems to eavesdrop on the key exchange and are realistically accessible to her. We show that under such scenarios a positive key is achievable. In a realistic scenario, it may be difficult for Eve to use a satellite to intercept the quantum signal between Alice and Bob, as Eve's satellite will need to match the relative speed of Alice's satellite, with respect to the OGS, in order to capture the entire signal exchange between Alice and Bob. Therefore, it may be more practical for Eve to use a HAP (or similar) to intercept the quantum signal between Alice and Bob. Using a HAP will simplify her ability to stay in between Alice and Bob, as HAP's are stationary with respect to a ground station \cite{aragon2008high}. The use of HAP as a relay station between satellite and OGS has been proposed for quantum communications \cite{vu2019hap,vu2020design} and for classical communications as well \cite{vu2018all,swaminathan2021haps}. HAP's are generally stationed at an altitude between $20~\text{km}$ and $10~\text{km}$ above sea level \cite{karapantazis2005broadband, kurt2021vision} for the best use case. Therefore, enabling Eve with a HAP may be a realistic assumption, if this sort of scheme is implemented in a real world scenario. Using this realistic restriction, it is assumed that Eve has the best channel and loss conditions. Therefore, assuming Eve is now closer to earth, she can be restricted further and more realistic parameters can be chosen with Alice and Bob still being able to generate a positive secret key.

\begin{figure}
\centering
\begin{subfigure}{0.5\textwidth}
\includegraphics[width=0.95\linewidth]{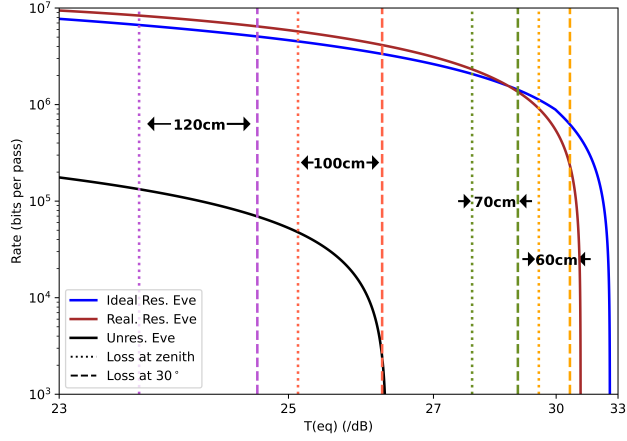}
\caption{}
\label{fig:SKR1a} 
\end{subfigure}
\medskip
\begin{subfigure}{0.5\textwidth}
\includegraphics[width=0.95\linewidth]{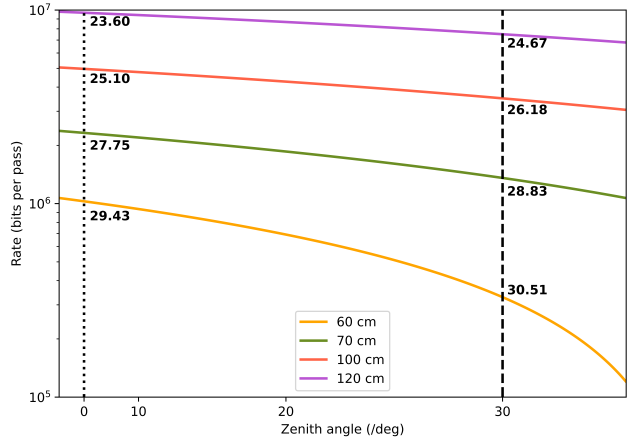}
\caption{}
\label{fig:SKR1b}
\end{subfigure} 
\caption{The achievable secret key rates per pass for the given parameter values as a function of channel loss (Fig. \ref{fig:SKR1a}) and Zenith angle (Fig. \ref{fig:SKR1b}) is plotted here. The coloured solid lines in Fig \ref{fig:SKR1a} indicate the key rate with restricted Eve assumption, key rate parameters given in Table \ref{tab:keyrate}. The black solid line indicates the conventional asymptotic secret key rate for an unrestricted Eve \cite{laudenbach2018continuous} with $V_{\text{opt}}=2.4~\text{SNU}$, $\beta=0.95$ and $\xi_{\text{tot}}=0.01$. The figure also provides information on possible key rates from the respective telescope aperture sizes, with dotted (dashed) line indicating the loss at zenith ($\pm30^{\circ}$). The secret key rate for realistic restricted Eve is shown as a function of Zenith angle for the various telescope aperture sizes in Fig. \ref{fig:SKR1b}. Here the assumptions and parameters are the same as in Fig \ref{fig:SKR1a}.}
\label{fig:SKR1}
\end{figure}

With the mutual information and restricted Eve's Helovo bound, a secret key can be calculated for the SPOQC CV-QKD payload. Here, the key rate calculation is limited to the asymptotic regime and is therefore a theoretical upper limit of the achievable secret key. In a realistic satellite-to-ground key exchange, the finite size of data will effect the secret key rate, as data will be used for parameter estimation. The true finite size effects are difficult to estimate, as the variance in the transmittance will effect accuracy of the parameter estimation. In order to overcome some of the data limitations due to repetition rates, satellite pass duration or interruptions due to weather conditions etc., multiple satellite passes can be used to generate a secret key of useful length. It is important to note that no positive secret key can be generated beyond $35~\text{dB}$ even given restricted Eve security assumptions, as the modulation variance cannot be increased beyond $300~\text{SNU}$ without leaking too much information to Eve, such that the Holevo bound is larger than the mutual information between Alice and Bob. For most receiver telescope apertures and wavelengths, CV-QKD will not be possible under adverse weather conditions due to the large increase in channel loss (see Tab. \ref{tab:weather} \& \ref{tab:weather_lambda}). Therefore, we will consider clear sky conditions when calculating the achievable secret keys for the CV-QKD payload in SPOQC mission and this is shown in Fig. \ref{fig:SKR1}. The figures show the secret key for receiver telescope apertures of $60~\text{cm}$ (orange), $70~\text{cm}$(green), $100~\text{cm}$ (red) and $120~\text{cm}$ (purple). This was done under ideal and realistic conditions as a function of $T_{eq}$ in Fig \ref{fig:SKR1a}. The figure shows the achievable secret key per pass for each of these receiver telescope aperture sizes at the zenith (dotted) and at zenith angle $\pm30^{\circ}$ (dashed). The parameters used to calculate the restricted Eve secret key rates can be seen in Tab. \ref{tab:keyrate}. The figure also shows the conventional fibre based asymptotic secret key rate. Here, the conventional asymptotic secret key rate can only achieve a positive rate around $26~\text{dB}$ channel loss, given an optimal modulation variance of $V_{\text{opt}}=2.4~\text{SNU}$, reconciliation efficiency of $\beta=0.95$ and excess noise of $\xi_{\text{tot}}=0.01$. This figure shows a positive secret key, for the CV-QKD payload in SPOQC mission, if restricted Eve security assumptions are made. It is important to note that the modulation variance used to calculate the secret key is optimised for expected channel loss Fig. \ref{fig:changing_ar}. With reduced channel loss, the modulation variance can be decreased, this means with increased receiver telescope aperture, the modulation variance can be decreased. When the receiver telescope aperture is doubled, the modulation variance can be decreased by an order of magnitude. In fibre based CV-QKD implementations the modulation variance is commonly a few SNU to tens of SNU \cite{zhang2020long}. The modulation variance used to calculate the key bits per pass are within the order of magnitude of fibre based implementations. Here, the bits per pass is given by $f_\text{rep} \times t_{\text{sat}} \times K$, where $f_\text{rep} = 2~\text{MHz}$ is the repetition rate and $t_{\text{sat}}=1.2~\text{s}$ is the ideal satellite overpass time. The bits per pass under realistic restricted Eve assumptions as a function of zenith angle for the various aperture sizes is shown in Fig. \ref{fig:SKR1b}. Here the same parameters and assumptions made in Fig. \ref{fig:SKR1a} were used, except the $V_{opt}$ was optimised specifically for each telescope aperture size.

\begin{widetext}

\begin{table}
\begin{center}
\begin{tabular}{||c | c | c | c | c ||} 
\hline
~Parameter ~&~ Blue line in Fig. \ref{fig:SKR1a} ~&~ Brown line in Fig. \ref{fig:SKR1a} ~&~ Fig. \ref{fig:SKR2a} ~&~ Fig. \ref{fig:SKR2b} ~\\ [0.5ex] 
\hline\hline
$V_{\text{opt}}$ & $12~\text{SNU}$ & $25~\text{SNU}$ & $300~\text{SNU}$ & $300~\text{SNU}$\\ 
\hline
$\eta_{AE}$ & $0.01$ & $0.005$ & $0.03$ & $0.03$ \\ 
\hline
$\eta_{T}$ & $0.1$ & $0.1$ & $0.4$ & $0.4$ \\
\hline
$\xi_{\text{tot}}$ & $0.001$ & $0.005$ & $0.001$ & $0.003$ \\ 
\hline
$\beta$ & $1$ & $0.96$ & $1$ & $1$ \\ 
\hline
$\eta_{d}$ & $1$ & $0.8$ & $1$ & $1$ \\ 
\hline
\end{tabular}
\caption{Parameters taken for plotting Fig. \ref{fig:SKR1} \& \ref{fig:SKR2} in addition to $v_{\text{el}}=0$ and $\eta_{S}=0$ being the same in them.}
\label{tab:keyrate}
\end{center}
\end{table}

\end{widetext}

The expected secret key per pass, for the SPOQC mission CV-QKD payload, given different signal wavelengths can be seen in Fig \ref{fig:SKR2}. In this figure the variances in secret key per pass for various wavelengths is shown at the zenith and at zenith angle $\pm30^{\circ}$. Fig \ref{fig:SKR2a} shows the achievable secret keys for nighttime operations and Fig. \ref{fig:SKR2b} for daytime operations. In order to show this variance non-deceptively, the figures are not in the conventional log scale used to display secret key rates. This figure shows an increase in variance of the secret key for shorter wavelengths, this is due to the increased variance in transmissivity for shorter wavelengths. The increased excess noise due to the variance in transmissivity has not been taken into account for these secret key calculations. Instead, the variance in secret key per pass is shown to highlight the increased stability of a pass at, given the choice of a larger wavelength such as $1550~\text{nm}$. The issue of increased excess noise due to the variance in transmissivity, particularly the large variance associated with shorter wavelengths, is a complex issue to resolve in post processing, and is not within the scope of this work. Instead, $1550~\text{nm}$ was chosen for the SPOQC mission CV-QKD signal wavelength, in order to circumvent this potential issue.

\section{Conclusion}
\label{conclusion}

A detailed outline of downlink based CV-QKD is given in the manuscript, with particular emphasis on the SPOQC mission. Various losses affect the performance in a space-based quantum communication scheme. Divergence is the primary factor that contributes to the maximum loss, along with beam wandering, turbulence \& scintillation, coupling losses, pointing \& tracking losses, and others. Here we characterise the quantum channel for all the potential losses in the case of the SPOQC mission CV-QKD protocol. We show that for a satellite in LEO, the potential loss could be $24-34~\text{dB}$ depending on the wavelength and telescope aperture areas (see Fig. \ref{fig:changing_ar}), given a clear-sky channel assumption is made. If cloud cover is considered, the channel loss is too large for secure key generation regardless of receiver telescope aperture size or signal wavelength (see Tab. \ref{tab:weather} \& \ref{tab:weather_lambda}). Instead, site diversity must be implemented to overcome the issue of cloud cover. Although, secret key generation is possible for partial cloud cover with a reduced key rate as the partial cloud cover will reduce the usable data for parameter estimation. When fog is considered, secret key generation is possible only in moderate fog conditions. It may be more practical to overcome all adverse weather conditions with OGS site diversity in satellite scheduling, this ensures the highest secret key rate during any given satellite pass.

In this work, various different turbulence strengths were considered for the loss characterisation. For $1550~\text{nm}$, low and medium turbulence have a negligible impact on the overall loss. Even in the high turbulence regime, the loss contributed by turbulence for $1550~\text{nm}$ is very small and secret key generation for the CV-QKD payload in SPOQC mission is possible even in the presence of strong turbulence. For wavelengths shorter than $1550~\text{nm}$, the impact of turbulence on channel loss will increase (Sec. \ref{tot_loss}). In a realistic scenario, the turbulence will most commonly be in the low to medium turbulence range \cite{walters1981atmospheric}. 

The choice of wavelength will have a significant impact on the losses the quantum signal experiences through the channel. As loss due to diffraction is the major contributor to the overall loss, the loss will increase significantly with increasing wavelength, as seen in Fig. \ref{fig:oss_vs_lambd}. Although with decreasing wavelength, the dynamic losses such as the one due to turbulence and pointing error would increase. The variance of these dynamic losses would also increase with decreasing wavelength. This implies that larger wavelengths will experience more overall loss, but their loss will be more stable compared with shorter wavelengths. With increased loss variance, parameter estimation may be more difficult, as in CV-QKD post processing parameters are estimated w.r.t the transmittance of the channel \cite{laudenbach2018continuous}. As a consequence, the longer wavelengths like $1550~\text{nm}$, may have less post processing overhead but will experience more overall loss. Although not within the scope of this work, one can optimise the wavelength for overall loss and loss variance to maximise the achievable secret key rate. For the SPOQC mission CV-QKD protocol, the $1550~\text{nm}$ signal wavelength was chosen to reduce the variance in loss, minimise background noise photons, and capitalise on the availability of high-efficiency, low-noise photo detectors and optical modulators at this wavelength.

Another consideration for space-based CV-QKD is the use of a LLO. The advantage of such an LO system is the ability to go to higher repetition rates \cite{soh2015self,qi2015generating}. When two separate lasers are used for a CV-QKD space-based protocol, as is the case for an LLO based scheme, the quantum signal will be shifted in frequency w.r.t. the LO due to the relative motion of the satellite to the OGS. This means the quantum signal will have a relative motion w.r.t. the LO laser and there will be a frequency shift between the two due to the Doppler effect. Therefore, a major draw back of an LLO in space-based CV-QKD is the Doppler effect, as satellite will be moving relative to the OGS \cite{schlake2025pulse, wang2021feasibility}. This is not an issue with TLO based scheme, which is considered in this work, as the quantum signal and LO will pass through the same channel, and there will be no relative motion between the two. Time of arrival differences between signal and LLO will be an issue for the clock, although this clock issue is not considered in this work and it is an interesting problem that can be addressed in future work. LLO based schemes will have increased loss and excess noise. Loss due to mode mismatching from wavefront disruption to the signal beam induced by scintillation and difference in time of arrival from the Doppler effect. Excess noise due to phase error induced to the signal beam from turbulence and frequency shifting from the Doppler effect. The turbulence based effects may be overcome with the use of adaptive optics. Although it is not possible to fully correct these errors, adaptive optics will significantly increase the complexity of the receiver set-up. To circumvent the potential issues from an LLO based implementation, a TLO scheme was chosen for the CV-QKD payload in SPOQC mission, which also removes the need for adaptive optics.

Though all of the channel losses are equally important, the major contributing factor in space-based CV-QKD is diffraction as shown in the results (see Fig. \ref{fig:oss_vs_lambd}), with contributions up-to 95\% for a downlink channel. Larger aperture telescopes (transmitting and receiving) can reduce the loss from diffraction in addition to the selection for optimal wavelength, signal variance or photon number per pulse for maximising the achievable secret key rate. Additional security assumptions may play a role along with novel OGS concepts such as array optical telescopes. Using larger telescopes are not without its own limitations, as with a larger receiver aperture, the fluctuations in loss will increase due to scintillation, Eq. (\ref{eqn:sigI}) and (\ref{eq:Aputure_av}), and may have an impact on the parameter estimation. Development in data post processing is necessary to address these issues however is not in the scope this work. Decreasing the satellite altitude also reduces diffraction-induced loss. Conversely, this lower altitude intensifies the effect of atmospheric turbulence on the channel because the signal beam's smaller diameter results in a higher interaction with the turbulent atmospheric layers. Consequently, the variance in signal loss is increased.

From the channel assumptions made and the channel parameters discussed in this work, a positive secret key can be generated from the CV-QKD payload in SPOQC even during daytime operations and in presence of strong turbulence as given in Fig. \ref{fig:SKR1} \& \ref{fig:SKR2}, with restricted Eavesdropping assumptions.

\section*{Acknowledgments}
The authors thank Dr. Masoud Ghalaii for the useful discussions. The authors acknowledge the funding support from EPSRC Quantum Communications Hub (Grant number EP/T001011/1) and from EPSRC Integrated Quantum Networks Hub (Grant number EP/Z533208/1). E.T.H.M thanks the School of Physics, Engineering and Technology, University of York for the PhD funding.

\bibliography{sample}
\appendix
\section{Displacement loss derivation} \label{appendix}
\begin{figure}
\centering
\includegraphics[width=0.8\linewidth]{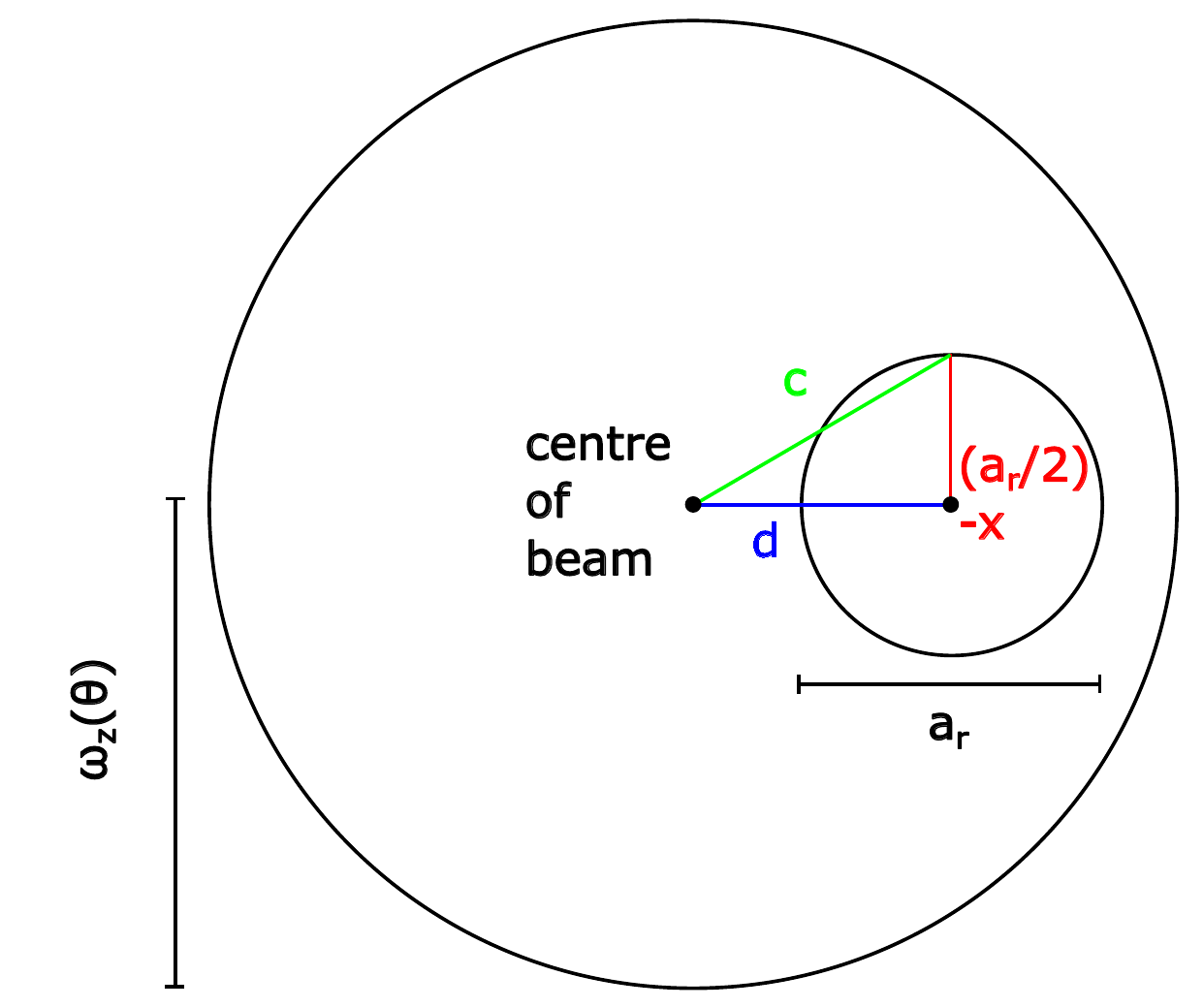}
\caption{The geometry used to derive the integrals used determine the loss for receiver apertures displaced from the centre of the transmitted beam. Here $c$ will be integral limit of the inner integrals, $x$ is the variable the outer integral integrates across and $d$ is the radial distance the receiver aperture is displace from the centre of the propagating beam. The receiver aperture is $a_r$ and the propagating beam width at the receiver is $\omega_z(\theta)$.}
\label{fig:appendix}
\end{figure}

We derive the loss for receiver apertures, displaced from the centre of the transmitted beam as shown in Eq. (\ref{eqn:T_dis}). To determine the transmittance from intensity, we must integrate all the intensities over the receiver area. From the Gaussian intensity profile given in Sec. \ref{channel_loss}, the intensity profile as a function the radial distance from the centre of the beam \cite{bea1991fundamentals} as,

\begin{equation}
I(r) = \frac{2P}{\pi \omega (\theta)^2}\exp\bigg(\frac{-2r^2}{\omega(\theta)^2}\bigg),
\end{equation}

where $r$ is the radial distance from the centre of the beam, $P$ is the power of the beam and $\omega(\theta)$ is the propagating beam width. Therefore, first we must determine the radial distances from the centre of the beam and use these values as our integral limits. From Fig. \ref{fig:appendix}, we can determine our integral limits using trigonometry. We will establish two integrals in order to integrate over the full receiver area (we will integrate over half the circle area and then multiply by two), with $c$ being the integral limit of the inner integral and $x$ as the variable the outer integral integrates across the radius. This means the outer integral will integrate along the line marked with red of length $a_r/2$ in the figure. The outer integral simply integrates across the radius of the receiver aperture (from $0$ to $a_r/2$). The second integral will run parallel along the length $d$ (the radial distance the receiver aperture is displace from the centre of the propagating beam). To determine the limits of this second integral using Pythagoras theorem, we find that

\begin{equation}
c=\sqrt{d^2+((a_r/2)-x)^2}.
\end{equation}

To determine the limits of the integral from this, we need to work out how far this limit is from $c$. Since we are integrating over a half circle, this will range from $x=0$ to $x=a_r/2$. Therefore, we subtract $x$ from $c$ for the lower limit and add $x$ to $c$ for the upper limit of this integral. Such that the lower limit $c_1=\sqrt{d^2+((a_r/2)-x)^2}-x$ and the upper limit $c_2=\sqrt{d^2+((a_r/2)-x)^2}+x$. Using these limits established and the Gaussian intensity profile, we integrate the intensities across the area of a half circle and multiply this by two to obtain the transmittance for any given $d$ and $a_r$. The integral we obtain is Eq. (\ref{eqn:T_dis}), with the respective notations given there.

\section{Photon flux and excess noise for free space detectors \label{app:flux}}

In free-space continuous-variable quantum key distribution (CV-QKD), background radiation collected by the receiver contributes to excess noise and limits system performance. In this section, we model the daytime and nighttime background photon flux and derive the corresponding excess noise contribution.

The fundamental physical constants used are
$h = 6.626 \times 10^{-34}\ \mathrm{J\,s}, ~c = 3.0 \times 10^{8}\ \mathrm{ms}^{-1}.$ The telescope aperture is assumed to be circular with with diameter $D = 2R = 70~\text{cm}$, corresponding to an effective collecting area $A_r \approx 0.269~\text{m}^2 (= \pi R^2)$. The CV receiver aperture is having the diameter $d = 2r = 3~\text{mm}$, corresponding to an effective collecting area $A \approx 7.06 \times 10^{-6}~\text{m}^2$. The telescope is assumed to have a receiving field of $\theta_f = 10\mu ~\mathrm{rad},$ having the field of view characterized by a solid angle $\Omega = 3.14 \times 10^{-10}\ \mathrm{Sr},$ the optical linewidth is $\Delta \lambda = 1~\mu\mathrm{m}$, and the gating window of the detector $\Delta t = 2~\mathrm{ns}$. The receiver temperature is taken to be $T = 300~\mathrm{K}$. Thus the total mode number of the noise photons which can be detected by the detector is \cite{qi2010feasibility}, 
\begin{align}
N_{mod} &= \Delta\nu \Delta t, \nonumber \\
&= \frac{c \Delta\lambda \Delta t}{\lambda^2}.
\end{align}

For an optical wavelength $\lambda$, the photon energy is $E_{\mathrm{ph}}(\lambda) = \frac{hc}{\lambda}$. We consider three wavelengths commonly used in free-space optical and QKD systems: $\lambda \in \{630,\,1064,\,1550\}\ \mathrm{nm}$ for all in $f(\lambda) \in \{\lambda_1,\,\lambda_2,\,\lambda_3\}$.

Given telescope parameters and brightness of the sky (wavelength dependent), the noise at the receiver is found as follows. The power received by telescope $P_b$ is \cite{er2005background, bonato2009feasibility},
\begin{align}
P_b = S_b \times \Omega \times A_r \times B_f,
\end{align}
where $S_b$ is the brightness of the background sky and $B_f = 1~\mathrm{nm}$ being the telescope filter bandwidth. The typical values of $S_b$ are given in Tab. \ref{tab:noise}.

\begin{table}[h]
\begin{center}
\begin{tabular}{||c | c | c | c ||} 
\hline
~Wavelength ~&~ $S_b$ - Clear ~&~ $S_b$ - Moonless ~& $N_{mod }$\\ [0.5ex] 
 (nm) & daytime & clear night & \\
\hline\hline
630 & 1.7 & $1.7\times10^{-5}$ &~ $1.5\times10^{6}$ ~\\ 
\hline
1064 & 0.9 & $0.9\times10^{-5}$ & $5.2\times10^{5}$ \\ 
\hline
1550 & 0.4 & $0.4\times10^{-5}$ & $2.4\times10^{5}$ \\
\hline
\end{tabular}
\caption{Parameters considered for plotting Fig. \ref{fig:noisevsflux}. $S_b$ is given in the units of $\mathrm{W}~\mathrm{m}^{-2}~\mathrm{Sr}^{-1}~\mu\mathrm{m}^{-1}$.}
\label{tab:noise}
\end{center}
\end{table}

Given the power received by the telescope, one can find the photon flux per pulse and photon flux per pulse per mode. Considering the repetition rate to be of $2~\mathrm{MHz}$, we find the number of photons per pulse during nighttime (daytime) to be of order $\approx10^{-6}$ ($\approx10^{-1}$) and number of photons per pulse per mode to be of order $\approx10^{-12}$ ($\approx 10^{-7}$), for 1550 nm. In CV-QKD systems, background photons contribute to excess noise, expressed in shot-noise units (SNU). The excess noise associated with a photon flux $\Phi$ is given by \cite{weedbrook2012gaussian}
\begin{equation}
\xi_{background} = \bigg(\frac{2\eta_d\,\Phi}{\,N_{mod}}\bigg)f_s,
\end{equation}
where $\eta_d ~(=0.8)$ is the detection efficiency and $f_s = 40~\mathrm{dB}$ is the scaling factor for estimating the error at Alice.

For daytime operation, the excess noise is
\begin{equation}
\boxed{\xi_{\mathrm{day}}(\lambda) = \frac{2\eta_d\,\Phi_{\mathrm{day}}(\lambda) ~f_s}{N_{mod}}}
\end{equation}
and, the excess noise for nighttime operation becomes
\begin{equation}
\boxed{\xi_{\mathrm{night}}(\lambda) = \frac{2\eta_d\,\Phi_{\mathrm{night}}(\lambda) ~f_s}{N_{mod}}}
\end{equation}

The results reveal several orders of magnitude difference between daytime and nighttime excess noise, driven primarily due to difference in brightness of the background sky. These effects strongly depend on wavelength, field of view, and detection bandwidth, highlighting the importance of careful system design for free-space CV-QKD links.

\section{Bypass channel covariance matrix and Holevo information} \label{apend:bypass}
This section of the Appendix shows the covariance matrix and Holevo bound for the bypass channel under the assumption of restricted Eve \cite{ghalaii2023satellite}. Here, the protocol is mapped to an entanglement based one (two mode squeezed vacuum state - TMSVS), to find the respective covariance matrices as well as Holevo information. And if $\eta_{S}$ can be set to zero, this effectively removes the bypass channel and gives Eve the optimal attack. The total covariance matrix is given by,

\begin{equation}
V_{ABEE^{\prime}}=\begin{pmatrix}
V\identity_2 & C_{AB}\mathbb{Z} & 0\identity_2 & C_{AE^{\prime}}\mathbb{Z}\\
C_{AB}\mathbb{Z} & V_{B}\identity_2 & C_{BE}\mathbb{Z} & C_{BE^{\prime}}\identity_2\\
0\identity_2 & C_{BE}\mathbb{Z} & V_{E}\identity_2 & C_{EE^{\prime}}\mathbb{Z}\\
C_{AE^{\prime}}\mathbb{Z} & C_{BE^{\prime}}\identity_2 & C_{EE^{\prime}}\mathbb{Z} & V_{E^{\prime}}\identity_2
\end{pmatrix},
\label{eqn:covarience}
\end{equation}

where mode $A$ and its covariances is the first row and column, mode $B$ in the second, mode $E$ in the third and mode $E^{\prime}$ in the fourth. Here modes $AB$ represent the TMSVS created by Alice and mode $EE^{\prime}$ being the TMSVS of Eve. The respective matrix elements in Eq. (\ref{eqn:covarience}) are defined as,

\begin{equation}
\begin{aligned}
C_{AB} &=\sqrt{T_{eq}}c,\\
C_{AE^{\prime}}&=-\sqrt{\eta_{AE}(1-\eta_{E})}c,\\
V_{B}&=T_{eq}(V-1)+1+\xi_{eq}^{Rx},\\
C_{BE}&=\sqrt{(1-\eta_{E})\eta_{T}}c_{e},\\
C_{BE^{\prime}}&=\sqrt{\eta_{E}(1-\eta_{E})\eta_{T}}(-(\eta_{AE}(V-1)+1)+V_{E})\\
&-\sqrt{\eta_{AE}(1-\eta_{AE})(1-\eta_{E})\eta_{S}(1-\eta_{T})}(V-1),\\
C_{EE^{\prime}}&=\sqrt{\eta_{E}}c_{E},\\
V_{E^{\prime}}&=(1-\eta_{E})[\eta_{AE}(V-1)+1]+\eta_{E}V_{E}.
\end{aligned}
\end{equation}

with,
\begin{equation}
\begin{aligned}
c&=\sqrt{V^{2}-1},\\
c_{E}&=\sqrt{V_{E}^{2}-1},\\
T_{eq}&=(\sqrt{\eta_{AE}\eta_{E}\eta_{T}}+\sqrt{(1-\eta_{AE})\eta_{S}(1-\eta_{T})})^{2},\\
\xi_{eq}^{Rx}&=T_{eq}\xi_{\text{tot}}=(1-\eta_{E})\eta_{T}(V_{E}-1),
\end{aligned}
\end{equation}

where $T_{eq}$ is the effective transmittance and $\xi_{eq}^{Rx}$ is the effective observed excess noise. $V$ is the modulation variance, $V_{E}$ is Eve variance, $\mathbb{Z}=diag(1,-1)$ and $\identity_2$ is the identity matrix. All other parameters have been defined previously. From the covariance matrix, the Holevo bound can be calculated. The Holevo information for reverse reconciliation is given by,
\begin{equation}
\chi_{BE}=H(EE^{\prime})-H(EE^{\prime}|B).
\end{equation}

Here $H(EE^{\prime})$ and $H(EE^{\prime}|B)$ can be obtained by the corresponding symplectic eigenvalues of the covariance matrix Eq. (\ref{eqn:covarience}) (for $EE^{\prime}$ and $EE^{\prime}|B$). And for $V_{EE^{\prime}}$, modes $A$ an $B$ have been traced out of the covariance matrix Eq. (\ref{eqn:covarience}) and the symplectic eigenvalues are denoted by $\Lambda_{1}$ and $\Lambda_{2}$. $V_{EE^{\prime}|B}$ can be obtained by applying a a homodyne measurement to mode B of the covariance matrix such that,
\begin{equation}
V_{EE^{\prime}|B}=V_{EE^{\prime}}-\frac{1}{V_{B}}\Sigma_{BEE^{\prime}}\Pi\Sigma^{T}_{BEE^{\prime}},
\end{equation}
where $\Sigma^{T}_{BEE^{\prime}}=\begin{bmatrix}
C_{BE}\mathbb{Z} & C_{BE^{\prime}}\identity_2
\end{bmatrix}$, and $\Pi=diag(1,0)$. $V_{EE^{\prime}|B}$ symplectic eigenvalues are denoted by $\Lambda_{3}$ and $\Lambda_{4}$. Such that the Holevo information is now given by,
\begin{equation}
\chi_{BE}=g(\Lambda_{1})+g(\Lambda_{2})-g(\Lambda_{3})-g(\Lambda_{4}),
\end{equation}
where $g(x)=(\frac{x+1}{2})\log_{2}(\frac{x+1}{2})-(\frac{x+1}{2})-\log_{2}(\frac{x+1}{2})$.

\end{document}